\documentclass[12pt]{article}
\usepackage[left=2.5cm,top=2.50cm,right=2.5cm,bottom=2.50cm]{geometry}
\usepackage{mathrsfs}
\usepackage{amsmath,amssymb,latexsym,color,cancel,graphicx,bbm,colortbl}
\usepackage[english]{babel}
\usepackage[latin1]{inputenc}
\usepackage{ragged2e}
\usepackage{float}
\usepackage{cite}
\usepackage{graphicx}
\usepackage{placeins}
\usepackage{nccmath}
\usepackage[toc,title]{appendix}
\usepackage{url}
\DeclareMathAlphabet{\mathpzc}{OT1}{pzc}{m}{it}
\begin{document}
\date{}
\title{Exact Solutions of the Dunkl--Pauli Equation in the Presence of a Magnetic Field from an $\mathfrak{su}(1,1)$ Algebraic Approach}
 \maketitle
\begin{center}
{\large M. Salazar Ram\'{i}rez$^{1}$}\\[0.3em]
\begin{minipage}{0.9\textwidth}
\centering
\small Escuela Superior de C\'omputo, Instituto Polit\'ecnico Nacional,\\
Av. Juan de Dios B\'atiz esq. Av. Miguel Oth\'on de Mendiz\'abal, Col. Lindavista,\\
Alc. Gustavo A. Madero, C.P. 07738, Ciudad de M\'exico, M\'exico.
\end{minipage}
\footnotetext[1]{msalazarra@ipn.mx(Corresponding author)}
\end{center}

\vspace{0.7em}
\begin{center}
{\large B. C. L\"utf\"uo\u{g}lu$^{2}$}\\[0.3em]

\begin{minipage}{0.9\textwidth}
\centering
\small Department of Physics, University of Hradec Kr\'alov\'e,\\
Rokitansk\'eho 62, 500 03 Hradec Kr\'alov\'e, Czechia.
\end{minipage}
\footnotetext[2]{bekir.lutfuoglu@uhk.cz}
\end{center}

\vspace{0.7em}
\begin{center}
{\large H. Bouguerne$^{3}$}\\[0.3em]

\begin{minipage}{0.9\textwidth}
\centering
\small Laboratoire de syst\`emes dynamiques et contr\^ole (L.S.D.C),\\
D\'epartement des sciences de la mati\`ere,\\
Facult\'e des Sciences Exactes et SNV,\\
Universit\'e de Oum-El-Bouaghi, 04000, Oum El Bouaghi, Algeria.
\end{minipage}
\footnotetext[3]{hacene.bouguerne@univ-oeb.dz}
\end{center}

\begin{abstract}
We investigate the two-dimensional Dunkl--Pauli equation for a spin--$1/2$ particle in the presence of an external magnetic field within an algebraic framework. While previous studies have provided exact solutions of the Dunkl--Pauli system, its underlying dynamical symmetry and algebraic structure remain largely unexplored. By employing Schr\"odinger factorization, we construct the generators of the $\mathfrak{su}(1,1)$ algebra and establish the associated symmetry of the radial sector. The energy spectrum is derived using irreducible unitary representations, and the corresponding Sturmian radial basis is obtained analytically. Furthermore, $\mathrm{SU}(1,1)$ coherent states are constructed and their time evolution is analyzed, revealing a characteristic radial breathing behavior. The results show that the Dunkl deformation introduces parity-dependent modifications in the spatial structure of the system while preserving its underlying algebraic dynamics.
\end{abstract}

\noindent\textbf{Keywords:} Dunkl--Pauli equation; Dunkl operators; Pauli equation;
$\mathfrak{su}(1,1)$ algebra; coherent states; magnetic field.

\section{Introduction}
The extension of quantum mechanics beyond the standard canonical framework has attracted considerable attention, particularly in the presence of discrete symmetries such as reflections. Early developments in this direction can be traced back to Wigner's proposal of modified commutation relations~\cite{Wigner1950} and Yang's construction of momentum operators incorporating reflection symmetries in the harmonic oscillator~\cite{Yang1951}. These ideas were later explored within deformed Heisenberg algebras, where reflection-deformed structures appeared in generalized quantization schemes leading to parafields and parastatistics~\cite{Green1953,Greenberg1964}, and were applied to systems such as the Calogero model~\cite{Brzezinski1993}, bosonization of supersymmetric quantum mechanics~\cite{Ply96}, anyonic systems~\cite{Ply94}, and more generally to deformations associated with generalized statistics~\cite{Ply97}.

A concrete operator realization of these concepts was established through the introduction of Dunkl operators, originally proposed by Dunkl~\cite{Dunkl1989}, which combine differential and reflection operators into a unified differential--difference structure. Within this formalism, the canonical momentum is modified through the replacement of spatial derivatives by Dunkl operators, leading to a consistent incorporation of discrete symmetries at the dynamical level. This construction forms the basis of Wigner--Dunkl quantum mechanics~\cite{Chung2019}, where the underlying algebra is deformed and the resulting dynamics depends explicitly on parity sectors~\cite{Dong22, Junker23, Chung23, Benchika24, Benchika25}. This operator-based formulation provides a natural framework for analyzing quantum systems in which symmetry and solvability are deeply intertwined, and has led to a broad range of exactly solvable models and applications~\cite{Genest2013, Genest2014, Genest2013b, Quesne24, HamilDdim2025}, including Dunkl oscillator systems and their nonrelativistic and relativistic extensions within algebraic frameworks~\cite{Salazar2017, Salazar2018, Mota2021, Ghaz}.

The applicability of the Dunkl formalism extends to relativistic quantum systems, where reflection operators lead to modifications of both spectral and dynamical properties. In particular, relativistic wave equations such as the Klein--Gordon equation have been formulated within the Dunkl framework, yielding exact analytical solutions for oscillator and Coulomb-type interactions~\cite{Hamil2022a}. Relativistic Dunkl oscillators have also been investigated in various settings, including Dirac-type systems in the presence of external magnetic fields, providing further insight into the structure of deformed wave equations~\cite{Mota2019}. In addition, thermodynamic aspects of relativistic Dunkl systems have been explored, where Klein--Gordon and Dirac oscillator formulations can be treated within a unified framework, allowing the evaluation of physical quantities such as mean energy, entropy, and specific heat~\cite{Hamil2022b, Rouabhia2023}. Furthermore, algebraic approaches based on $\mathfrak{su}(1,1)$ symmetry have been employed to obtain exact spectra, eigenfunctions, and coherent states, highlighting the role of symmetry methods in the analysis of relativistic Dunkl systems~\cite{Salazar2026}.

In nonrelativistic quantum mechanics, Dunkl operators allow the exact or approximate treatment of a broad class of potentials, including harmonic~\cite{Dong2023, Hamil2}, pseudoharmonic~\cite{Mota2022}, Coulomb~\cite{Genest2015,Ghazouani2020}, P\"oschl--Teller~\cite{SchulzeHalberg2024}, Morse~\cite{Hamil2025Morse}, and more general anharmonic interactions~\cite{SchulzeHalberg2024b}. These studies show that the deformation parameters modify the spectral structure and influence physical observables through the presence of reflection operators.

An important extension of this framework arises in spin-dependent systems, where the coupling between spin and external magnetic fields introduces additional degrees of freedom. Within this context, the Dunkl--Pauli equation has been formulated and solved for spin--$1/2$ particles, yielding parity-dependent energy spectra and spinor eigenfunctions~\cite{Bouguerne2024}. Time-dependent generalizations, such as the Dunkl--Pauli oscillator, further show that the inclusion of Dunkl derivatives significantly modifies the dynamical evolution of quantum states~\cite{Benchikha2025}. Recent studies have also explored thermodynamic and time-dependent aspects of Dunkl--Pauli systems in the presence of Aharonov--Bohm flux, highlighting the continued development of this framework~\cite{Tedjani2026,Khantoul2026}.

Despite these developments, the algebraic structure underlying the Dunkl--Pauli system has not been fully characterized. In particular, while exact solutions are available, an explicit construction of the associated $\mathfrak{su}(1,1)$ dynamical symmetry, together with the derivation of the corresponding Sturmian basis and coherent states, remains largely unexplored from an algebraic perspective. Without this algebraic framework, the organization of the spectrum and the description of quantum-state dynamics cannot be systematically addressed.

In this work, we investigate the two-dimensional Dunkl--Pauli equation for a spin--$1/2$ particle in a uniform magnetic field from an algebraic perspective. By employing Schr\"odinger factorization, we construct the generators of the $\mathfrak{su}(1,1)$ algebra and establish the corresponding dynamical symmetry of the system. This framework enables us to derive the energy spectrum and eigenfunctions, as well as to obtain the associated Sturmian radial basis in a systematic manner. We further construct the $\mathrm{SU}(1,1)$ coherent states and analyze their time evolution. Our approach provides a unified algebraic formulation of the Dunkl--Pauli system and elucidates the role of reflection symmetry in shaping its spectral and dynamical properties.

The $\mathfrak{su}(1,1)$ algebra encodes the infinitesimal structure of the system and underlies the Lie group $\mathrm{SU}(1,1)$ through exponentiation, yielding finite transformations. Within this framework, the algebra provides a natural setting for the construction of ladder operators and the factorization of the Hamiltonian, while the associated group plays a central role in the formulation of coherent states via displacement operators and in their time evolution \cite{Wybourne,Hall}.

This paper is organized as follows. In Sec.~2, we present the theoretical framework of the two-dimensional Dunkl--Pauli equation. In Secs.~3 and 4, we develop the $\mathfrak{su}(1,1)$ algebraic treatment and obtain the corresponding physical quantities, together with the associated graphical analysis, for the case of symmetric Dunkl parameters $(|\nu_1|=|\nu_2|)$. Subsequently, in Sec.~5, we extend this analysis to the case of asymmetric Dunkl parameters $(|\nu_1|\neq |\nu_2|)$ by presenting the corresponding plots and comparing both behaviors. Finally, in Sec.~6, we summarize the main results and discuss their physical implications.
\section{A Review of  Dunkl--Pauli Equation}
The stationary Pauli equation, obtained from the nonrelativistic limit of the Dirac theory while retaining the spin--magnetic interaction, can be written as\cite{Bouguerne2024}
\begin{equation}
\frac{1}{2m}(\vec{\pi}\!\cdot\!\vec{\sigma})^{2}\psi = E\psi,
\label{PE}
\end{equation}
where the conjugate momenta include the electromagnetic coupling through $\pi_{j}=p_{j}-\tfrac{e}{c}A_{j}$.
In the Dunkl framework, the canonical momenta are replaced by differential-reflection operators, introducing a discrete symmetry into the kinematic structure\cite{Dunkl1989}. They are defined as
\begin{equation}\label{OPRE}
p_{j}=-iD_{j},\qquad D_{j}=\partial_{x_{j}}+\frac{\nu_{j}}{x_{j}}(1-R_{j}),
\end{equation}
here, $D_j$ denotes the Dunkl differential--reflection operator, $\nu_j$ are the deformation parameters, and $R_j$ represent reflection operators associated with the coordinate $x_j$. The reflection operators introduce a parity-dependent structure in the system, meaning that the dynamics explicitly depends on the behavior of the wavefunction under spatial inversion.

It is worth emphasizing that the operators defined in equation \ref{OPRE} correspond to the reflection group $W=\mathbb{Z}_2 \times \mathbb{Z}_2$, associated with independent reflections in each spatial coordinate. In a more general setting, Dunkl operators can be constructed for arbitrary finite reflection groups $W$, leading to richer algebraic structures \cite{Feigin,DeBie}. Their construction relies on operators satisfying
\begin{equation}
R_{j}f(x_{j})=f(-x_{j}),\qquad R_{j}R_{i}=R_{i}R_{j},\qquad R_{j}x_{i}=-\delta_{ij}x_{i}R_{j}.
\label{reflection_ops}
\end{equation}
These operators satisfy the commutation relations
\begin{equation}
[x_{i},D_{j}]=\delta_{ij}(1+2\nu_{j}R_{j}),\qquad [D_{i},D_{j}]=0,\qquad [x_{i},x_{j}]=0.
\label{eq8}
\end{equation}

This structure is characteristic of Dunkl--type algebras and may be viewed as a realization of a rational Cherednik algebra associated with the reflection group $W$. Unlike a simple Heisenberg deformation, the inclusion of reflection operators introduces parity-dependent terms that enhance the symmetry structure of the system \cite{Etingof,LePage,Hakobyan}.

The Wigner parameters are assumed to satisfy $\nu_j>-1/2$ which guarantees a well-defined weighted Hilbert space $\mathcal{H}\!\left(\mathbb{R}^2,|x_1|^{2\nu_1}|x_2|^{2\nu_2}dx_1dx_2\right)$ and the integrability of the associated Dunkl measure. In this framework, the scalar product is naturally defined with the weighted measure $|x_1|^{2\nu_1}|x_2|^{2\nu_2}dx_1dx_2$, under which the Dunkl momentum operators remain Hermitian and the reflection structure of the system is consistently incorporated\cite{Chung2019,Junker2024}. A uniform external magnetic field is introduced through the symmetric gauge choice $\vec{A}=\tfrac{B}{2}(-x_{2}\hat{i}+x_{1}\hat{j})$, under which the Dunkl--Pauli Hamiltonian takes the form \cite{Bouguerne2024}
\begin{equation}
\medmath{H =-\frac{1}{2m}\frac{\partial^{2}}{\partial r^{2}}-\frac{1+2\nu_{1}+2\nu_{2}}{2mr}\frac{\partial}{\partial r}+\frac{m\omega_{c}^{2}}{8}\,r^{2}
+\frac{\mathcal{B}_{\theta}}{mr^{2}}+\frac{\omega_{c}}{2}\mathcal{J}_{\theta}- g_{s}\mu_{B}\left(1+\nu_{1}R_{1}+\nu_{2}R_{2}\right)\boldsymbol{B\cdot{S}},
\label{eq:DPH_polar}}
\end{equation}
where $\omega_c = eB/(mc)$ and $\mu_B = |e|/(2mc)$ denote the cyclotron frequency and the Bohr magneton, respectively, while $\boldsymbol{B}$ represents the external magnetic field strength. The spin operator is given by $\mathbf{S}=\boldsymbol{\sigma}/2$ and $g_s=2.0023$ is the free electron $g$-factor. The operators $B_\theta$ and $J_\theta$ represent the angular contributions within the Dunkl framework.

Thus, the angular dependence of the Dunkl deformation is encoded in the operators\cite{Bouguerne2024}
\begin{align}\label{ECBT}
\mathcal{B}_{\theta} &=-\frac{1}{2}\partial_{\theta}^{2}+(\nu_{1}\tan\theta-\nu_{2}\cot\theta)\partial_{\theta}+ \frac{\nu_{1}}{2\cos^{2}\theta}(1-R_{1})
+\frac{\nu_{2}}{2\sin^{2}\theta}(1-R_{2}),\\
\mathcal{J}_{\theta} &=i\!\left[\partial_{\theta}+\nu_{2}\cot\theta(1-R_{2})-\nu_{1}\tan\theta(1-R_{1})\right].
\label{ECAT}
\end{align}
Because $\mathcal{J}_{\theta}$ commutes with $R_{1}R_{2}$, the Hilbert space splits into reflection sectors, and the wavefunction separates as\cite{Genest2013,Genest2014}
\begin{equation}
\phi_{\epsilon,m_{s}}(r,\theta)=F_{m_{s}}(r)\,\Theta_{\epsilon}(\theta),
\label{eq24}
\end{equation}
and the angular factor satisfies the eigenvalue equation
\begin{equation}
\mathcal{J}_{\theta}\Theta_{\epsilon}(\theta)=\lambda_{\epsilon}\Theta_{\epsilon}(\theta),
\label{eq25}
\end{equation}
here $\lambda_{\epsilon}$ is the eigenvalue of Dunkl-angular operator. For $\epsilon=+1$, this case corresponds to the situation in which either
$\epsilon_{1}=\epsilon_{2}=+1$ or $\epsilon_{1}=\epsilon_{2}=-1$(eigenvalues of the reflection operators $R_1$ and $R_2$, respectively); therefore, the angular solutions take the form\cite{Bouguerne2024}
\begin{equation}
\Theta_{+1}(\theta)=a_{\ell}\,\mathbf{P}_{\ell}^{(\nu_{1}+1/2,\nu_{2}+1/2)}(-2\cos\theta)\pm a_{\ell}'\,\sin\theta\cos\theta\,\mathbf{P}_{\ell-1}^{(\nu_{1}+1/2,\nu_{2}+1/2)}(-2\cos\theta),
\label{eq26}
\end{equation}
with
\begin{equation}\label{LAM1}
\lambda_{+}=\pm 2\sqrt{\ell(\ell+\nu_{1}+\nu_{2})},\qquad\ell\in\mathbb{N}^{*}.
\end{equation}
Here, $\mathbf{P}_{\ell-1}^{(\alpha,\beta)}(x)$ denote Jacobi polynomials, obtained via the change of variable $x=-2\cos\theta$ with $\theta\in[0,\pi]$ ($x\in[-2,2]$), and $\ell$ corresponds to the angular momentum of the system. By identifying the radial function in the separated wavefunction as $F_{m_s}(r)=F^{(\epsilon_1,\epsilon_2)}_{m_s}(r)$, the radial component of the Dunkl--Pauli equation assumes the general form\cite{Bouguerne2024}
\begin{equation}\label{ECDSO}
\medmath{
\left[\frac{d^{2}}{dr^{2}}+\frac{1+2\nu_{1}+2\nu_{2}}{r}\frac{d}{dr}-\frac{m^{2}\omega_{c}^{2}}{4}r^{2}-\frac{\lambda_{+}^{2}}{r^{2}}-m\omega_{c}\lambda_{+}
+ mB\mu_{B}g_{s}m_{s}(1+\nu_{1}\epsilon_{1}+\nu_{2}\epsilon_{2})+2mE\right]F_{\ell,m_s}^{\epsilon_{1},\epsilon_{2}}(r)=0.}
\end{equation}
The radial equation shows that the Dunkl parameters $\nu_1$ and $\nu_2$ modify the effective centrifugal term and the spin-magnetic interaction, thereby affecting both the localization and energy spectrum of the system.

\section{Schr\"odinger factorization and $\mathfrak{su}(1,1)$ structure: Case $\boldsymbol{\epsilon=+1}$ ($\epsilon_1=\epsilon_2=1$) and ($|\nu_1|=|\nu_2|$)}
The use of Schr\"odinger factorization allows us to reveal an underlying algebraic structure, which provides a systematic way to obtain the energy spectrum and eigenfunctions. By introducing the dimensionless radial variable $\rho =\sqrt{\frac{m\omega_c}{2}}\, r$, the radial equation (\ref{ECDSO}) can be cast into the dimensionless form
\begin{equation}\label{EDSOPP}
\medmath{\left[\frac{d^{2}}{d\rho^{2}}+\frac{1+2\nu_{1}+2\nu_{2}}{\rho}\frac{d}{d\rho}-\rho^{2}-\frac{\lambda_{+}^{2}}{\rho^{2}}-2\lambda_{+}+\frac{2 B \mu_{B} g_{s} m_{s}}{\omega_{c}}
\left(1+\nu_{1}\epsilon_{1}+\nu_{2}\epsilon_{2}\right)+\frac{4E}{\omega_{c}}\right]\phi_{\ell,m_s}^{\epsilon_{1},\epsilon_{2}}(\rho)=0}.
\end{equation}
To eliminate the first-order derivative term appearing in the dimensionless radial equation, we introduce the transformation
\begin{equation}\label{SecondC}
\phi_{\ell,m_s}^{\epsilon_{1},\epsilon_{2}}(\rho)=\frac{\Psi_{\ell,m_s}^{\epsilon_{1},\epsilon_{2}}(\rho)}{\rho^{\gamma}},
\end{equation}
where $\gamma=\frac{1}{2}\left(1+2\nu_{1}+2\nu_{2}\right)$. The first and second derivatives associated with this transformation are respectively given by

\begin{align}
\frac{d\phi_{\ell,m_s}^{\epsilon_{1},\epsilon_{2}}(\rho)}{d\rho}=&\left[-\gamma\rho^{-(\gamma+1)}+\rho^{-\gamma}\frac{d}{d\rho}\right]\Psi_{\ell,m_s}^{\epsilon_{1},\epsilon_{2}}(\rho),\\
\frac{d^{2}\phi_{\ell,m_s}^{\epsilon_{1},\epsilon_{2}}(\rho)}{d\rho^{2}}=&\left[\gamma(\gamma+1)\rho^{-(\gamma+2)}-2\gamma \rho^{-(\gamma+1)}\frac{d}{d\rho}+\rho^{-\gamma}\frac{d^{2}}{d\rho^{2}},
\right]\Psi_{\ell,m_s}^{\epsilon_{1},\epsilon_{2}}(\rho).
\end{align}
Substituting these expressions into the radial equation (\ref{EDSOPP}) and simplifying the resulting terms, the first-order derivative contribution is completely removed.

Consequently, after multiplying by $-\rho^2$, the equation for $\Psi_{\ell,m_s}^{\epsilon_{1},\epsilon_{2}}(\rho)$ assumes a Schr\"odinger--like form. In particular, for the sector $\epsilon_1=\epsilon_2=1$, one obtains
\begin{equation}\label{edso2}
\medmath{\left[-\rho^{2}\frac{d^{2}}{d\rho^{2}}+\rho^{4}-\tfrac{1}{4}+\lambda_{+}^{2}+(\nu_{1}+\nu_{2})^{2}+\left(2\lambda_{+}-\frac{2 B \mu_{B} g_{s} m_{s}}{\omega_{c}}
\left(1+\nu_{1}+\nu_{2}\right)-\frac{4E}{\omega_{c}}\right)\rho^2\right]\Psi_{\ell,m_s}^{+,+}(\rho)=0 },
\end{equation}
within the Schr\"odinger factorization formalism, a closed set of $\mathfrak{su}(1,1)$ generators is constructed by adopting the ansatz
\begin{equation}\label{FACS}
\left[\rho\frac{d}{d\rho}+\mathcal{A}\rho^2+\mathcal{B}\right]\biggl[-\rho\frac{d}{d\rho}+\mathcal{C}\rho^2+\mathcal{F}\biggr]\Psi_{\ell,m_s}^{+,+}(\rho)=\mathcal{G}\Psi_{\ell,m_s}^{+,+}(\rho),
\end{equation}
expanding the factorization ansatz in Eq.~(\ref{FACS}) and collecting terms with equal powers of $\rho$, one obtains the second--order differential equation
\begin{equation}\label{EDSOEX}
\medmath{\left[-\rho^{2}\frac{d^{2}}{d\rho^{2}}-\rho\left((-\mathcal{C}+\mathcal{A})\rho^{2}+\mathcal{B}+1-\mathcal{F}\right)\frac{d}{d\rho}
+\mathcal{A}\mathcal{C}\rho^{4}+\left((\mathcal{B}+2)\mathcal{C}+\mathcal{A}\mathcal{F}\right)\rho^{2}+\mathcal{B}\mathcal{F}\right]\Psi_{\ell,m_s}^{+,+}(\rho)
=\mathcal{G}\Psi_{\ell,m_s}^{+,+}(\rho)}.
\end{equation}
A direct term-by-term comparison between Eq.~(\ref{EDSOEX}) and the Schr\"odinger--like equation~(\ref{edso2}) yields the following system of algebraic relations for the factorization parameters
\begin{align}\nonumber
\mathcal{A}-\mathcal{C}=0,\quad\mathcal{B}+1-\mathcal{F}=&0,\quad\mathcal{A}\mathcal{C}=1,\quad \mathcal{G}-\mathcal{B}\mathcal{F}=\frac{1}{4}-\lambda_{+}^{2}-(\nu_{1}+\nu_{2})^{2},\\
(\mathcal{B}+2)\mathcal{C}+\mathcal{A}\mathcal{F}=&2\lambda_{+}-\frac{2 B \mu_{B} g_{s} m_{s}}{\omega_{c}}\left(1+\nu_{1}+\nu_{2}\right)-\frac{4E}{\omega_{c}},
\end{align}
resulting in the factorization parameters $\mathcal{A}$, $\mathcal{B}$, $\mathcal{C}$, $\mathcal{F}$, and $\mathcal{G}$ listed in Table~1.
\begin{table}[H]
\centering
\caption{Schr\"odinger factorization parameters for the sector $\epsilon_{1}=\epsilon_{2}=1$.}
\label{tab:ABCFGmm1}
\begin{tabular}{ccccc}
\hline\hline
$\mathcal{A}$ & $\mathcal{B}$ & $\mathcal{C}$ & $\mathcal{F}$ & $\mathcal{G}$ \\
\hline
$+1$ & $-\chi^{+,+}-\tfrac{3}{2}$ & $+1$ & $-\chi^{+,+}-\tfrac{1}{2}$
& $\lambda_{+}^{2}+(\nu_{1}+\nu_{2})^{2}+(\chi^{+,+}+1)^{2}-\frac{1}{2}$ \\[1mm]
$-1$ & $+\chi^{+,+}-\tfrac{3}{2}$ & $-1$ & $+\chi^{+,+}-\tfrac{1}{2}$
& $\lambda_{+}^{2}+(\nu_{1}+\nu_{2})^{2}+(\chi^{+,+}-1)^{2}-\frac{1}{2}$ \\
\hline\hline
\end{tabular}
\end{table}
where $\displaystyle \chi^{+,+} = -\lambda_{+}+\frac{B\mu_{B} g_{s} m_{s}}{\omega_{c}} \left(1+\nu_{1}+\nu_{2}\right)+\frac{2E}{\omega_{c}}$.

Table~$1$ shows the values of the parameters $\mathcal{A}$, $\mathcal{B}$, $\mathcal{C}$, $\mathcal{F}$ and $\mathcal{G}$ that appear implicitly in Eq.~(\ref{FACS}), obtained by expanding this equation and comparing
it term by term with Eq.~(\ref{edso2}). In this case, the conditions correspond to $\epsilon=+1$ ($\epsilon_1=\epsilon_2=1$). The separation constant is given by $\lambda_{+}=\pm2\sqrt{\ell(\ell+\nu_1+\nu_2)}$, with $\ell\in\mathbb{N}^{*}$.

It then follows that the equation satisfied by $\Psi_{\ell,m_s}^{+,+}(\rho)$ can be cast in the factorized form
\begin{equation}
\left( \Gamma_{\mp}^{+,+}\mp 1\right)\Gamma_{\pm}^{+,+}=\frac{1}{4}\left[-\lambda_{+}^{2}-(\nu_{1}+\nu_{2})^{2}+(\chi^{+,+}\pm 1)^{2}\right],
\end{equation}
where
\begin{equation}
\Gamma_{\mp}^{+,+}=\frac{1}{2}\left(\rho\frac{d}{d\rho}+\rho^{2}-\chi^{+,+}\pm\frac{1}{2}\right).
\end{equation}
These results allow us to construct two new operators
\begin{align}\label{OPSB1}
\Pi_{\pm}^{+,+}=\frac{1}{2}\left(\mp\rho\frac{d}{d\rho}+\rho^{2}\pm\frac{1}{2}-2\Lambda_{0}^{+,+}\right),
\end{align}
here,
\begin{equation}\label{TEROPC1}
\Lambda_{0}^{+,+}\Psi_{\ell,m_s}^{+,+}(\rho)= \frac{1}{4}\left[-\frac{d^{2}}{d\rho^{2}}+\rho^{2}+\frac{\lambda^{2}+(\nu_{1}+\nu_{2})^{2}-\tfrac{1}{4}}{\rho^{2}}\right]\Psi_{\ell,m_s}^{+,+}(\rho)=\frac{\chi^{+.+}}{2}\Psi_{\ell,m_s}^{+,+}(\rho),
\end{equation}
where $\chi^{+,+}/2$ corresponds to the eigenvalue associated with the operator $\Lambda_0^{+,+}$ acting on the radial Sturmian functions. Eqs.~(\ref{OPSB1}) and (\ref{TEROPC1}) are readily seen to define a closed realization of the $\mathfrak{su}(1,1)$ algebra (see Appendix for additional details)
\begin{equation}\label{TEROP3}
[\Lambda_{0}^{+,+},\Pi_{\pm}^{+,+}]=\mp\Pi_{\pm}^{+,+}, \hspace{0.5cm} [\Pi_{-}^{+,+}, \Pi_{+}^{+,+}]=2\Lambda_{0}^{+,+}.
\end{equation}
In order to extract the energy spectrum of the Dunkl--Pauli equation in the presence of a magnetic field, we make use of the unitary irreducible representations of the noncompact Lie algebra $\mathfrak{su}(1,1)$~\cite{Barut1,Perelomov}
\begin{equation}\label{OPTLA}
\Lambda_{0}^{+,+}|k, n\rangle=(k+n) |k,n\rangle,
\end{equation}
the quadratic Casimir operator $\mathscr{C}^2$ satisfies the eigenvalue relation
\begin{align}\label{CASC1}\nonumber
\mathscr{C}^2\Psi_{\ell,m_s}^{+,+}(\rho) =& (-\Pi_+^{+,+} \Pi_-^{+,+} +\Lambda_{0}^{+,+}(\Lambda_{0}^{+,+}-1))\Psi_{\ell,m_s}^{+,+}(\rho)=\frac{1}{4}\left[\lambda_+^2+\left(\nu_1+\nu_2\right)^2\right]\Psi_{\ell,m_s}^{+,+}(\rho)\\
=&k(k-1)\Psi_{\ell,m_s}^{+,+}(\rho),
\end{align}
the above equation determines the Bargmann index $k$ associated with the present problem in the form
\begin{equation}\label{INBAR}
k=\frac{1}{2} + \frac{1}{2}\sqrt{\lambda^{2}+\left(\nu_1+\nu_2\right)^2}=\frac{1}{2}+\frac{1}{2}\sqrt{4\ell\,(\ell+\nu_{1}+\nu_{2})+\left(\nu_1+\nu_2\right)^2}=\ell+\frac{1}{2}\left(\nu_1+\nu_2+1\right),
\end{equation}
the quantization condition follows immediately from the combined use of Eqs.~(\ref{TEROPC1}), (\ref{OPTLA}), and (\ref{INBAR})
\begin{equation}
 -\lambda_{+}+\frac{B\mu_{B} g_{s} m_{s}}{\omega_{c}} \left(1+\nu_{1}+\nu_{2}\right)+\frac{2E}{\omega_{c}}=2(n_r+\ell)+(1+\nu_1+\nu_2),
\end{equation}
from which the Dunkl--Pauli energy spectrum in a magnetic field is obtained
\begin{equation}
E^{+,+}_{n_r,\ell,m_s}=\omega_c\left[n_r+\ell+\frac{1+\nu_1+\nu_2}{2}+\sqrt{\ell(\ell+\nu_1+\nu_2)}-\frac{m_s}{2}(1+\nu_1+\nu_2)
\right].
\end{equation}
In the undeformed limit $\nu_1=\nu_2=0$, this expression reduces to
\begin{equation}\label{ENERNUC}
E^{+,+}_{n_r,\ell,m_s}=\omega_c\left[n_r+2\ell+\frac{1-m_s}{2}\right],
\end{equation}
which corresponds to the ordinary Pauli--Landau spectrum obtained when the Dunkl deformation and reflection-dependent contributions vanish.

The differential equation that allows one to construct the wave function $\Psi_{\ell,m_s}^{+,+}(\rho)$ corresponding to Eq.~(\ref{edso2}) is
\begin{equation}\label{eqdsol1}
y''+\left[4n_r+2\beta+2-x^2+\frac{\frac{1}{4}-\beta^2}{x^2}\right]y=0,
\end{equation}
which has the solution~\cite{Lebedev,Gur}
\begin{equation}\label{soldi2}
y=N_{n_r}e^{-\frac{x^2}{2}}x^{\alpha+\frac{1}{2}}L_{n_r}^{\alpha}\left(x^2\right).
\end{equation}
A direct use of Eqs.~(\ref{eqdsol1}) and (\ref{soldi2}) allows one to construct the wave functions $\Psi_{\ell,m_s}^{+,+}(\rho)$ corresponding to Eq.~(\ref{edso2}), which yields
\begin{align}\nonumber
\medmath{\Psi_{\ell,m_s}^{+,+}(\rho)} &=\medmath{\sqrt{\frac{2\,n_r!}{\Gamma\!\left(n_r+\sqrt{\lambda_{+}^{2}+(\nu_1+\nu_2)^{2}}+1\right)}}\;
e^{-\rho^{2}/2}\;\left(\frac{m\omega_c}{2}\right)^{\frac{1+\nu_1+\nu_2}{2}}\rho^{\,\sqrt{\lambda_{+}^{2}+(\nu_1+\nu_2)^{2}}+\frac{1}{2}}\;
L_{n_r}^{\,\sqrt{\lambda_{+}^{2}+(\nu_1+\nu_2)^{2}}}\!\left(\rho^{2}\right)} \\
&= \medmath{\sqrt{\frac{2\,n_r!}{\Gamma\!\left(n_r+2\ell+\nu_1+\nu_2+1\right)}}\;e^{-\rho^{2}/2}\;\left(\frac{m\omega_c}{2}\right)^{\frac{1+\nu_1+\nu_2}{2}}
\rho^{\,2\ell+\nu_1+\nu_2+\frac{1}{2}}\;L_{n_r}^{\,2\ell+\nu_1+\nu_2}\!\left(\rho^{2}\right)},
\end{align}
the irreducible unitary representation of the $\mathfrak{su}(1,1)$ Lie algebra for the Dunkl--Pauli equation in the presence of a magnetic field admits a Sturmian basis characterized by the group indices $n$ and $k$, of the form
\begin{align}
F_{n_r,\ell}^{+,+}(r)=&\sqrt{\frac{2\,n_r!}{\Gamma\!\left(n_r+2k\right)}}\;e^{-\frac{m\omega_c}{4}r^{2}}\;\left(\frac{m\omega_c}{2}\right)^{k}
r^{2\ell}\;L_{n_r}^{\,2k-1}\!\left(\frac{m\omega_c}{2}\,r^{2}\right).
\end{align}
\begin{figure}[H]
    \centering
    \includegraphics[width=0.90\textwidth]{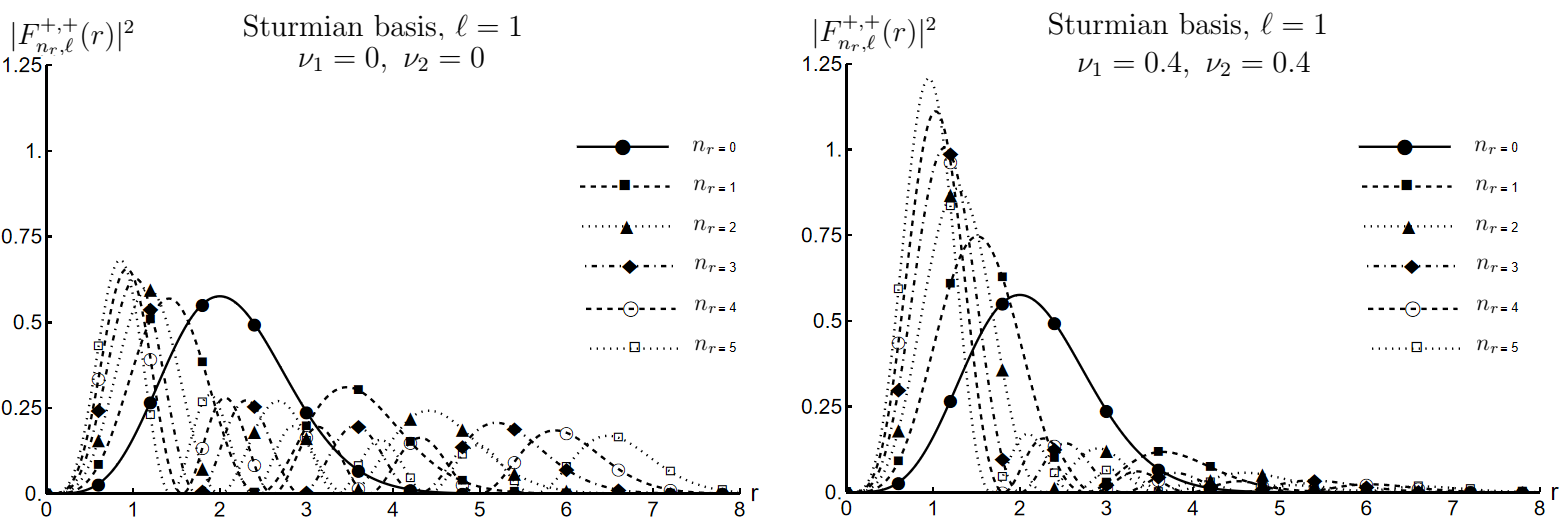}
    \caption{Normalized radial probability density $|F_{n_r,\ell}^{+,+}(r)|^2$
of the Sturmian basis for $\ell=1$, $(\nu_1,\nu_2)=(0,0)$ and $(0.4,0.4)$,
with $n_r=0,1,\ldots,5$.}
\end{figure}

\begin{figure}[H]
    \centering
    \includegraphics[width=0.90\textwidth]{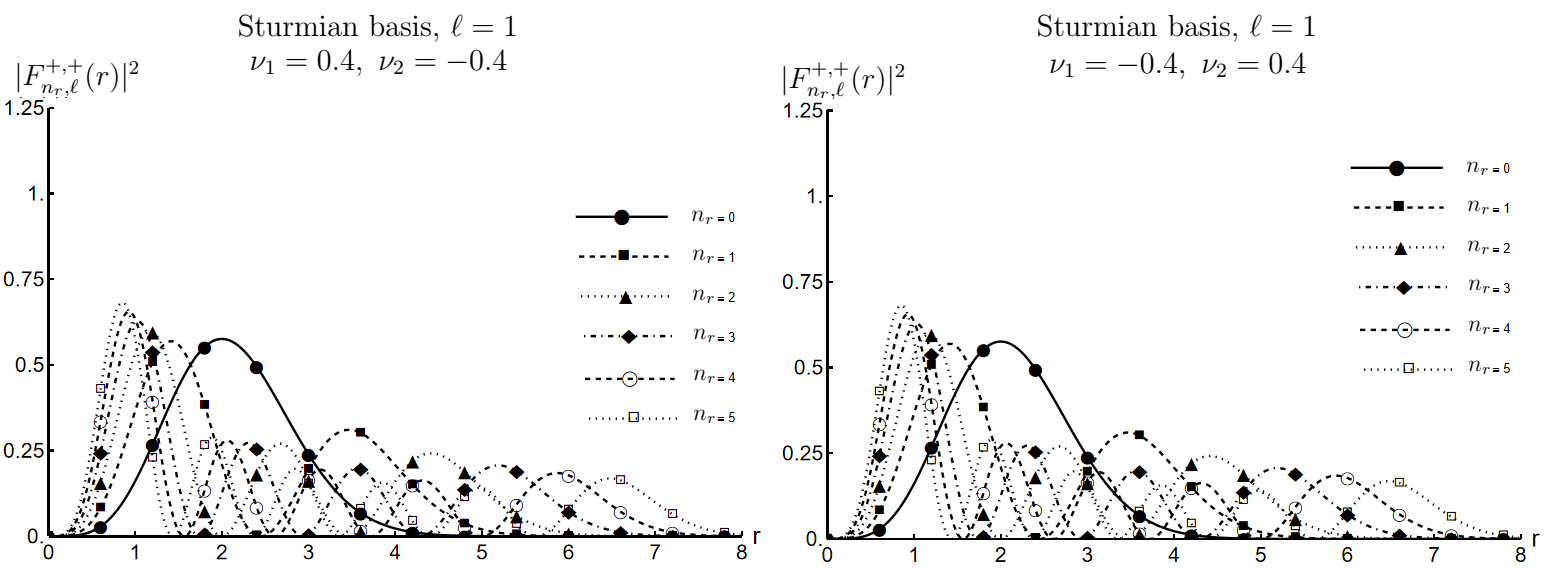}
    \caption{Normalized radial probability density $|F_{n_r,\ell}^{+,+}(r)|^2$
of the Sturmian basis for $\ell=1$, $(\nu_1,\nu_2)=(0.4,-0.4)$ and $(-0.4,0.4)$,
with $n_r=0,1,\ldots,5$.}
\end{figure}

\begin{figure}[H]
    \centering
    \includegraphics[width=0.50\textwidth]{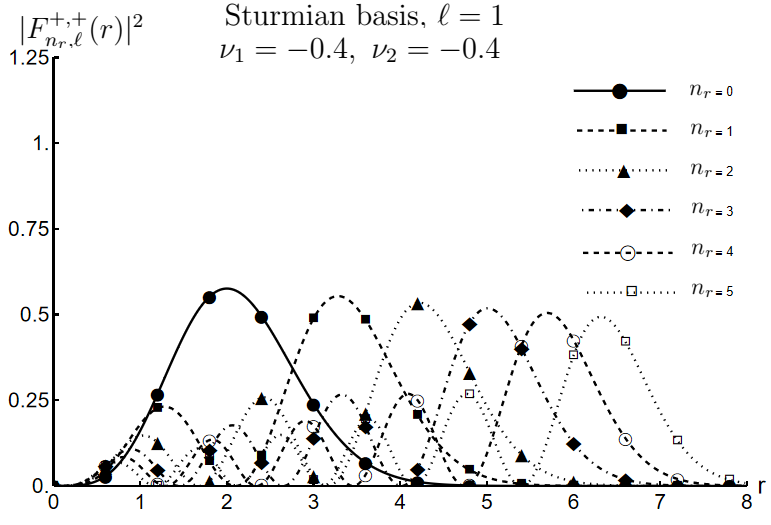}
    \caption{Normalized radial probability density $|F_{n_r,\ell}^{+,+}(r)|^2$
of the Sturmian basis for $\ell=1$, $(\nu_1,\nu_2)=(-0.4,-0.4)$ , with $n_r=0,1,\ldots,5$.}
\end{figure}

The radial probability densities $|F_{n_r,\ell}^{+,+}(r)|^2$ for $\ell=1$ and $n_r=0,\dots,5$ are displayed in Figs.~$1$--$3$ for different deformation sectors $(\nu_1,\nu_2)$. In all cases, as $n_r$ increases, the principal maximum shifts toward larger radial distances, i.e., to the right along the radial axis $r$, and an increasing number of nodes appears. A key observation is that the radial profiles depend solely on the combination $\nu_1 + \nu_2$; consequently, different pairs $(\nu_1, \nu_2)$ yielding the same sum produce identical radial probability densities.

The Dunkl deformation primarily modifies the position of these maxima: for $\nu_1+\nu_2>0$, the curves move to larger values of $r$ due to a stronger effective angular repulsion near the origin, whereas for $\nu_1+\nu_2=0$ they remain close to the undeformed case. In contrast, when $\nu_1+\nu_2<0$, the effective centrifugal contribution is reduced and the radial profile undergoes a global readjustment of its characteristic scale; however, the qualitative behavior under increasing $n_r$ remains unchanged, since the number of nodes grows and the principal maximum continues to shift toward larger radii. This behavior confirms that the effective parameter $k$ governs the degree of radial confinement of the states.

\subsection{$\mathrm{SU}(1,1)$ radial coherent states and their time evolution}
We define the $\mathrm{SU}(1,1)$ Perelomov coherent states as the action of the displacement operator on the lowest normalized state (see Appendix for additional details)
\begin{equation}
\label{eqnew}
|\zeta\rangle = D(\xi)\,|k,0\rangle=(1-|\xi|^{2})^{k}\sum_{n=0}^{\infty}\left[\frac{\Gamma(n+2k)}{n!\,\Gamma(2k)}\right]^{1/2}\xi^{n}\,|k,n\rangle ,
\end{equation}
applying the displacement operator to the ground radial state of the Dunkl oscillator, the coherent radial wave function can be written as
\begin{equation}
\label{eq38_new}
R(r,\xi)=\left[\frac{2(1-|\xi|^{2})^{2k}}{\Gamma(2k)}\right]^{1/2}\exp\left(-\frac{m\omega_c}{4}r^{2}\right) \left(\sqrt{\frac{m\omega_c}{2}} r\right)^{2k-\frac{1}{2}}
\sum_{n=0}^{\infty}\xi^{n}L^{2k-1}_{n}\left(\frac{m\omega_c}{2}r^{2}\right) .
\end{equation}
In order to evaluate the above series, we make use of the generating function of the Laguerre polynomials
\begin{equation}
\label{eq39_new}
\sum_{n=0}^{\infty}L^{\nu}_{n}(x)\,y^{n}=\frac{1}{(1-y)^{\nu+1}}\exp\!\left(-\frac{x\,y}{1-y}\right),\qquad |y|<1,
\end{equation}
using this identity, the summation in Eq.~(\ref{eq38_new}) can be carried out in closed form, yielding
\begin{equation}
\label{eq40_new}
R(r,\xi)= \left[\frac{2(1-|\xi|^{2})^{2k}}{\Gamma(2k)\,(1-\xi)^{4k}}\right]^{1/2}r^{\,2k-\frac{1}{2}}\exp\!\left[\frac{m\omega_cr^{2}}{2}\,\frac{\xi+1}{\xi-1}\right],
\end{equation}
finally, expressing the result in terms of the orbital quantum number
$\ell$ together with the Wigner parameters $\nu_{1}$ and $\nu_{2}$, we obtain the $\mathrm{SU}(1,1)$ radial coherent states for the Dunkl--Pauli equation in the presence of a magnetic field in the form
\begin{align}\label{COHE1}
\medmath{R(r,\xi)=\left[\frac{2(1-|\xi|^2)^{2\ell+\nu_1+\nu_2+1}}{\Gamma(2\ell+\nu_1+\nu_2+1)\,(1-\xi)^{2(2\ell+\nu_1+\nu_2+1)}}\right]^{\!1/2}\left(\frac{m\omega_c}{2}\right)^{\ell+\frac{\nu_1+\nu_2+1}{2}}r^{2\ell}\,
\exp\!\left[\frac{m\omega_c\,r^{2}}{4}\,\frac{\xi+1}{\xi-1}\right] .}
\end{align}
These coherent states describe semiclassical-like wave packets, whose evolution captures the dynamical behavior of the system in a compact analytical form.

The temporal evolution of these coherent states can be obtained in a direct way, since the third generator of the algebra is proportional to the radial Hamiltonian, namely $\Lambda_{0}^{+,+}=\tfrac{1}{2}H_{r}$. Consequently, the operator governing the time translation of the system is defined as (see Appendix for additional details)
\begin{equation}
\mathcal{T}(t)=e^{-i H_{r} t/\hbar}=e^{-i 2 \Lambda_{0} t/\hbar},
\label{Tt}
\end{equation}
where $t$ denotes an effective evolution parameter. Accordingly, the time-dependent Perelomov coherent state is introduced as
\begin{equation}
|\zeta(t)\rangle= \mathcal{T}(t)\,|\zeta\rangle= \mathcal{T}(t)\,D(\xi)\,\mathcal{T}^{\dagger}(t)\,\mathcal{T}(t)\,|k,0\rangle,
\label{zetat_def_T}
\end{equation}
from the action of the time-translation operator on the lowest-weight state, one immediately finds
\begin{equation}
\mathcal{T}(t)\,|k,0\rangle= e^{-2 i k t/\hbar}\,|k,0\rangle.
\label{evol_ground_T}
\end{equation}
The similarity transformation of the ladder operators follows from the Baker--Campbell--Hausdorff formula together with the $\mathfrak{su}(1,1)$ commutation relations, yielding
\begin{equation}\label{Piplus_T}
\Pi_{+}(t)= \mathcal{T}^{\dagger}(t)\,\Pi_{+}\,\mathcal{T}(t)= \Pi_{+}\,e^{2 i t/\hbar},\quad \Pi_{-}(t)=\mathcal{T}^{\dagger}(t)\,\Pi_{-}\,\mathcal{T}(t)= \Pi_{-}\,e^{-2 i t/\hbar}.
\end{equation}
As a consequence, the displacement operator evolves according to
\begin{equation}
\mathcal{T}(t)\,D(\xi)\,\mathcal{T}^{\dagger}(t)= \exp\!\left[\xi(-t)\Pi_{+}-\xi^{*}(-t)\Pi_{-}\right],
\label{D_T}
\end{equation}
where the time-dependent parameter is defined as $\xi(t)=\xi\,e^{2 i t/\hbar}$. In terms of this parameter, the normal-ordered form of the displacement operator reads
\begin{equation}
D(\xi(t))= \exp\!\left[\zeta(t)\Pi_{+}\right]\,\exp\!\left[\eta \Lambda_{0}\right]\,\exp\!\left[-\zeta^{*}(t)\Pi_{-}\right],
\label{D_normal_T}
\end{equation}
with $\zeta(t)=\zeta\,e^{2 i t/\hbar}$. Combining Eqs.~(\ref{evol_ground_T}) and (\ref{D_normal_T}), the time-evolved Perelomov coherent state takes the compact form
\begin{equation}
|\zeta(t)\rangle= e^{-2 i k t/\hbar}\,e^{\zeta(-t)\Pi_{+}}\,e^{\eta \Lambda_{0}}\,e^{-\zeta^{*}(-t)\Pi_{-}}\,|k,0\rangle,
\label{zetat_final_T}
\end{equation}
with these results, the configuration-space representation of the time-dependent radial coherent state becomes
\begin{equation}\label{EVTEC}
\medmath{
\begin{aligned}
R (r,\xi(t)) &=\left[\frac{2\bigl(1-|\xi|^{2}\bigr)^{2\ell+\nu_{1}+\nu_{2}+1}}{\Gamma\!\bigl(2\ell+\nu_{1}+\nu_{2}+1\bigr)\,\bigl(1-\xi e^{2it/\hbar}\bigr)^{2\left(2\ell+\nu_{1}+\nu_{2}+1\right)}}\right]^{\!1/2}\\[2mm]
&\times\exp\!\left[-2i\!\left(\ell+\frac{\nu_{1}+\nu_{2}+1}{2}\right)\frac{t}{\hbar}\right]\,\left(\frac{m\omega_{c}}{2}\right)^{\ell+\frac{\nu_{1}+\nu_{2}+1}{2}}\,r^{2\ell}\,\exp\!\biggl[\frac{m\omega_{c}r^{2}}{4}\,
\frac{\xi e^{2it/\hbar}+1}{\xi e^{2it/\hbar}-1}\biggr],\qquad |\xi|<1.
\end{aligned}}
\end{equation}

\begin{figure}[H]
    \centering
    \includegraphics[width=0.50\textwidth]{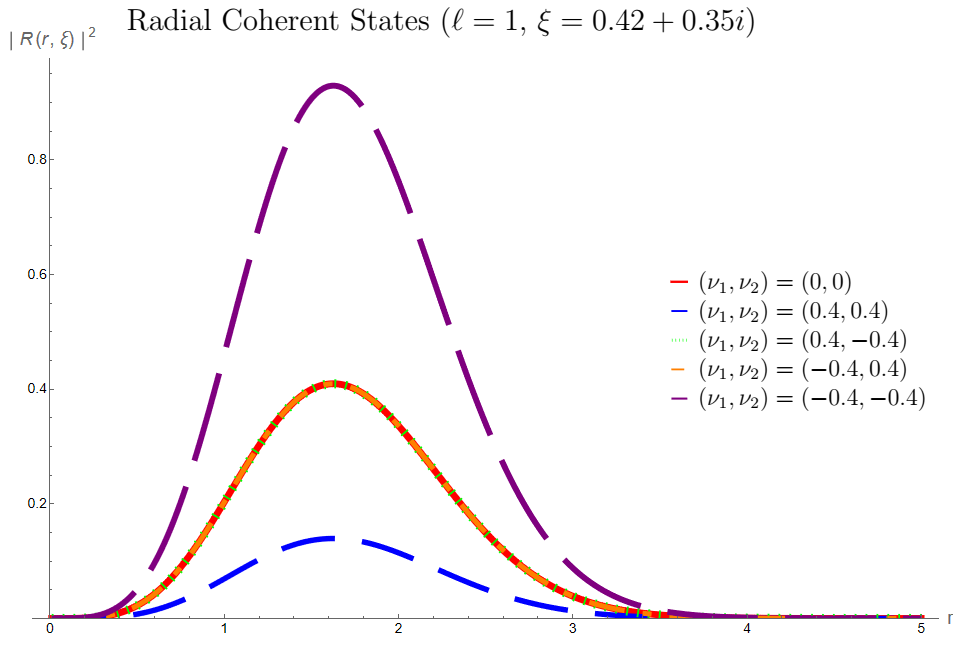}
    \caption{Radial coherent-state probability density $|R(r,\xi)|^2$ for $\ell=1$,
$\xi=0.42+0.35i$, and $(\nu_1,\nu_2)=(0,0),(0.4,0.4),(0.4,-0.4),(-0.4,0.4),(-0.4,-0.4)$.}
\end{figure}

\begin{figure}[H]
    \centering
    \includegraphics[width=0.50\textwidth]{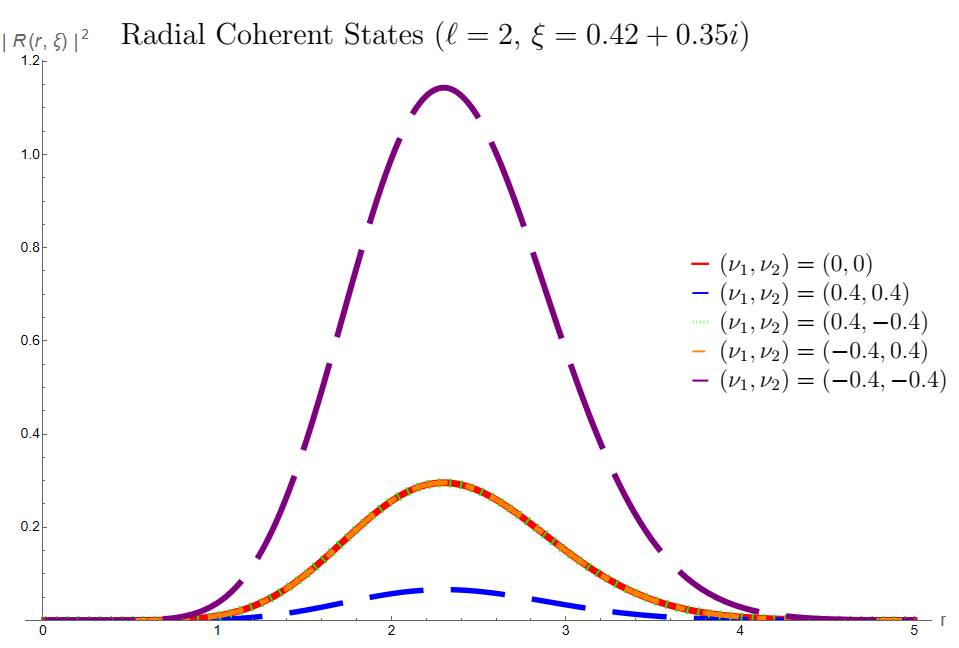}
    \caption{Radial coherent-state probability density $|R(r,\xi)|^2$ for $\ell=2$,
$\xi=0.42+0.35i$, and $(\nu_1,\nu_2)=(0,0),(0.4,0.4),(0.4,-0.4),(-0.4,0.4),(-0.4,-0.4)$.}
\end{figure}

\begin{figure}[H]
    \centering
    \includegraphics[width=0.50\textwidth]{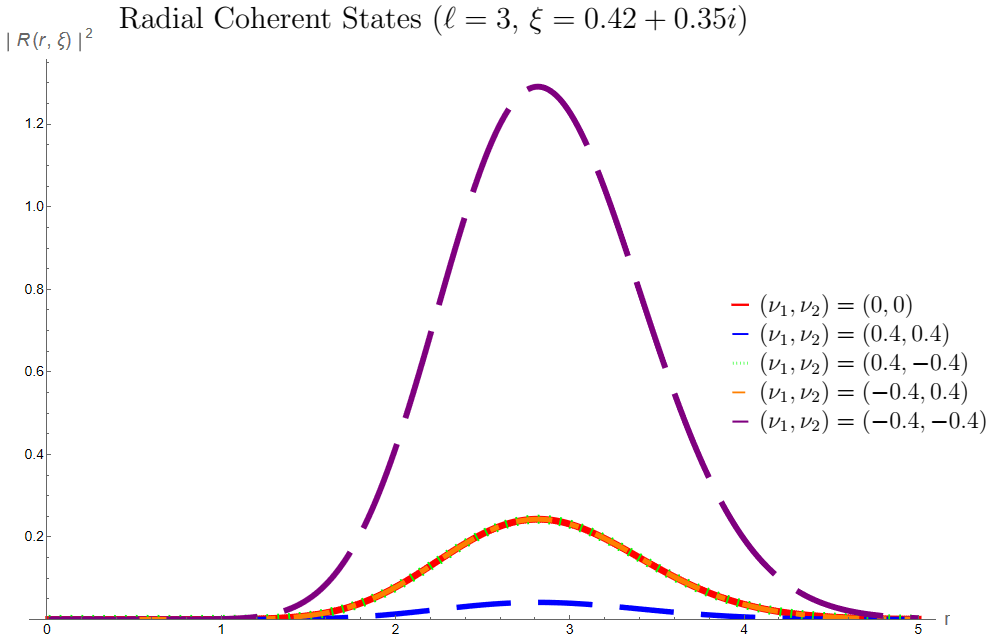}
    \caption{Radial coherent-state probability density $|R(r,\xi)|^2$ for $\ell=3$,
$\xi=0.42+0.35i$, and $(\nu_1,\nu_2)=(0,0),(0.4,0.4),(0.4,-0.4),(-0.4,0.4),(-0.4,-0.4)$.}
\end{figure}

\begin{figure}[H]
    \centering
    \includegraphics[width=0.50\textwidth]{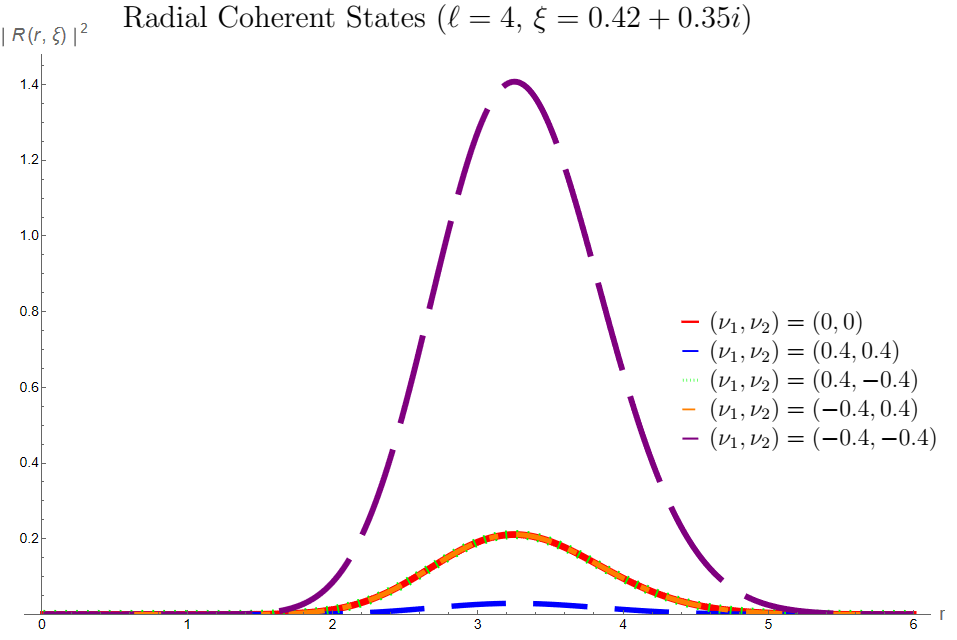}
    \caption{Radial coherent-state probability density $|R(r,\xi)|^2$ for $\ell=4$,
$\xi=0.42+0.35i$, and $(\nu_1,\nu_2)=(0,0),(0.4,0.4),(0.4,-0.4),(-0.4,0.4),(-0.4,-0.4)$.}
\end{figure}

\begin{figure}[H]
    \centering
    \includegraphics[width=0.50\textwidth]{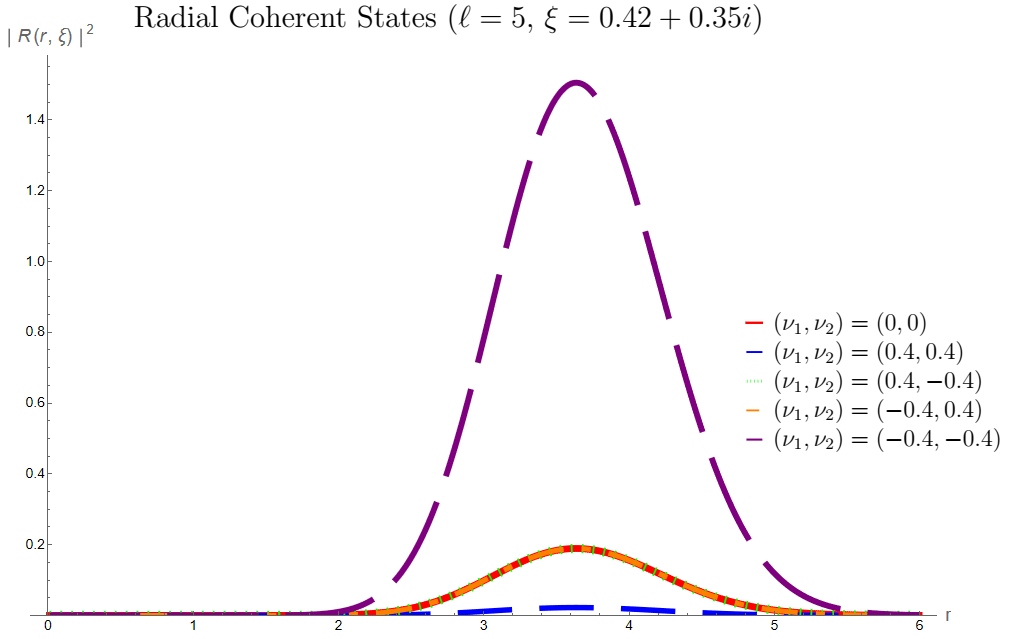}
    \caption{Radial coherent-state probability density $|R(r,\xi)|^2$ for $\ell=5$,
$\xi=0.42+0.35i$, and $(\nu_1,\nu_2)=(0,0),(0.4,0.4),(0.4,-0.4),(-0.4,0.4),(-0.4,-0.4)$.}
\end{figure}

As the orbital quantum number $\ell$ increases (Figs.~4--8), the radial probability density of the coherent state moves away from the origin and
progressively shifts toward larger radii. This behavior is characteristic of a coherent wave packet in the presence of an increasing centrifugal barrier:
the factor $r^{4\ell}$ suppresses the probability near $r=0$, while the exponential term preserves the overall confinement, producing a ring-like
distribution whose effective radius grows with $\ell$.

Consequently, the coherent state evolves from being concentrated near the center to describing
increasingly external radial orbits, analogous to the growth of the classical radius in a rotational oscillator. The overlapping curves correspond to
sectors $(\nu_1,\nu_2)$ with the same effective combination $\nu_1+\nu_2$, and therefore generate the same radial dynamics. Notably, all radial expressions depend only on the sum $\nu_1 + \nu_2$; consequently, pairs $(\nu_1,\nu_2)$ with the same sum yield identical curves.

\begin{figure}[H]
    \centering
    \includegraphics[width=0.96\textwidth]{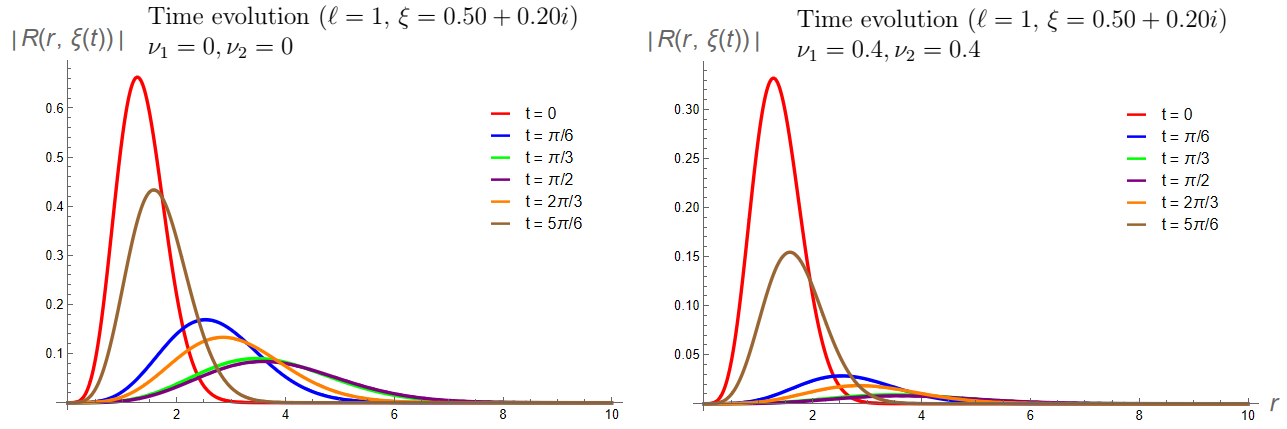}
    \caption{Time evolution of the radial coherent-state probability density $|R(r,\xi(t))|^{2}$ for $\ell=1$, $\xi=0.50+0.20i$, and $(\nu_1,\nu_2)=(0,0)$ and $(0.4,0.4)$ at $t=0,\pi/6,\pi/3,\pi/2,2\pi/3,5\pi/6$.}
\end{figure}

\begin{figure}[H]
    \centering
    \includegraphics[width=0.96\textwidth]{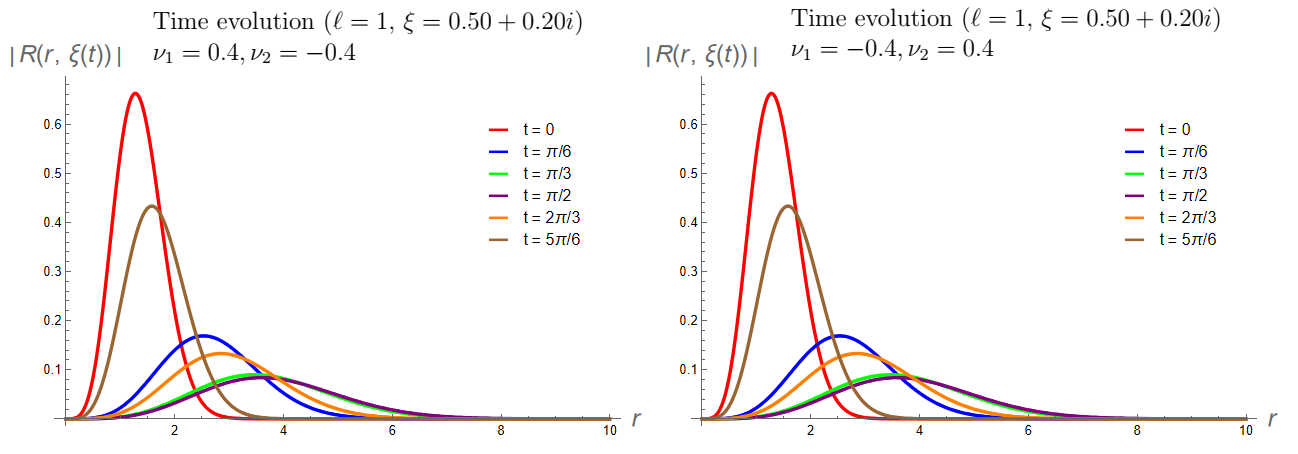}
    \caption{Time evolution of the radial coherent-state probability density $|R(r,\xi(t))|^{2}$ for $\ell=1$, $\xi=0.50+0.20i$, and $(\nu_1,\nu_2)=(0.4,-0.4)$ and $(-0.4,0.4)$ at $t=0,\pi/6,\pi/3,\pi/2,2\pi/3,5\pi/6$.}
\end{figure}

\begin{figure}[H]
    \centering
    \includegraphics[width=0.50\textwidth]{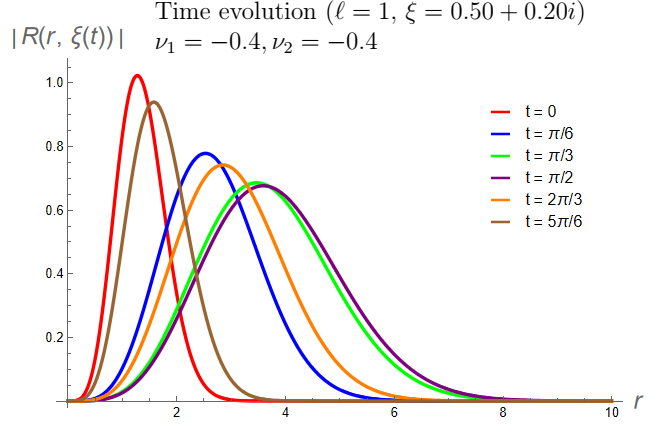}
    \caption{Time evolution of the radial coherent-state probability density $|R(r,\xi(t))|^{2}$ for $\ell=1$, $\xi=0.50+0.20i$, and $(\nu_1,\nu_2)=(-0.4,-0.4)$ at $t=0,\pi/6,\pi/3,\pi/2,2\pi/3,5\pi/6$.}
\end{figure}

Figs.~$9$--$11$ display the time evolution of the radial probability density $|R(r,\xi(t))|^{2}$ for $\ell=1$. Within each panel, each curve corresponds
to a fixed instant of time; thus, the dynamics is inferred by comparing the full temporal sequence. As $t$ increases, the probability maximum first
shifts toward larger values of $r$ and, as the sequence continues, appears again at positions closer to the origin within the periodic cycle,
revealing a radial ``breathing'' motion. During this process, the profile alternates between narrower and higher shapes and broader and lower ones,
reflecting the compression and expansion of the coherent packet.

Comparing different deformation sectors shows that the temporal pattern is common to all of them, changing only the radius around which the oscillation
occurs. When $\nu_1+\nu_2>0$, the entire evolution is shifted toward larger radii; when $\nu_1+\nu_2=0$, the curves remain practically coincident with
the undeformed case; and when $\nu_1+\nu_2<0$, the dynamics moves closer to the origin. Thus, the Dunkl deformation modifies the mean position of the
coherent packet without altering the periodicity or the nature of its temporal evolution, highlighting that all radial expressions depend solely on the sum $\nu_1 + \nu_2$.

\section{Schr\"odinger factorization and $\mathfrak{su}(1,1)$ structure Case: $(\epsilon_1,\epsilon_2)\in\{(-,-),(+,-),(-,+)\}$ and ($|\nu_1|=|\nu_2|$)}

In this section, we extend the algebraic treatment of the Dunkl--Pauli equation to the remaining reflection sectors. The analysis is carried out for the configurations $(\epsilon_1,\epsilon_2)\in\{(-,-),(+,-),(-,+)\}$, which together with the case $(\epsilon_1,\epsilon_2)=(+,+)$ studied in Sec.~3 correspond to $\epsilon=\pm 1$.

Although the radial equations differ among sectors, the underlying $\mathfrak{su}(1,1)$ algebraic structure remains the same. In particular, all cases share the same quadratic Casimir operator and Bargmann index, leading to identical Sturmian bases and coherent states. The differences arise only in the corresponding quantization conditions and energy spectra.

In what follows, we analyze each sector separately, emphasizing only the sector-dependent contributions while avoiding repetition of the algebraic construction.

\subsection{Case $\epsilon=+1$ $(\epsilon_1=\epsilon_2=-1)$}

In this sector, both reflection operators act with negative parity, which modifies the effective spin--magnetic coupling through the combination $(1-\nu_1-\nu_2)$. The corresponding radial equation reads
\begin{equation}
\medmath{\left[\frac{\partial^{2}}{\partial r^{2}}+\frac{1+2\nu_{1}+2\nu_{2}}{r}\frac{\partial}{\partial r}-\frac{m^{2}\omega_{c}^{2}}{4}r^{2}-\frac{\lambda_{+}^{2}}{r^{2}}-m\omega_{c}\lambda_{+}+mB\mu_{B}g_{s}m_{s}(1-\nu_{1}-\nu_{2})+2mE\right]F_{\ell,m_{s}}^{-,-}(r)=0.}
\end{equation}

After removing the first-order derivative term, the equation acquires a Schr\"odinger-like form
\begin{equation}
\medmath{\left[\frac{d^{2}}{d\rho^{2}}-\rho^{2}+\frac{\left[\lambda_{+}^{2}+(\nu_{1}+\nu_{2})^{2}\right]-\tfrac{1}{4}}{\rho^{2}}-2\lambda_{+}
+ \frac{2 B \mu_{B} g_{s} m_{s}}{\omega_{c}}\left(1-\nu_{1}-\nu_{2}\right)+\frac{4E}{\omega_{c}}\right]\Psi_{\ell,m_s}^{-,-}(\rho)=0.}
\end{equation}
The structure of the inverse-square term reflects the effective centrifugal barrier modified by the Dunkl parameters, while the constant terms encode the combined effect of the magnetic field and spin interaction.

By following a procedure analogous to that developed in Sec.~3 for the $(+,+)$ sector, as outlined in Eqs.~(19)--(25), one arrives at the quantization condition
\begin{equation}
-\lambda_{+}+\frac{B\mu_{B}g_{s}m_{s}}{\omega_{c}}(1-\nu_{1}-\nu_{2})+\frac{2E}{\omega_{c}}=2(n_{r}+\ell)+(1+\nu_{1}+\nu_{2}),
\end{equation}
from which the energy spectrum follows as
\begin{equation}
E^{-,-}_{n_r,\ell,m_s}=\omega_c
\left[n_r+\ell+\frac{1+\nu_1+\nu_2}{2}+\sqrt{\ell(\ell+\nu_1+\nu_2)}-\frac{m_s(1-\nu_1-\nu_2)}{2}\right].
\end{equation}
This result shows that the Dunkl deformation modifies the Zeeman-type splitting through the factor $(1-\nu_1-\nu_2)$, producing a sector-dependent shift in the energy levels.

\subsection{Case $\epsilon=-1$ $(\epsilon_1=1,\epsilon_2=-1)$}
In this mixed-parity sector, the reflection operators act asymmetrically, leading to a different effective coupling governed by $(\nu_1-\nu_2)$. The radial equation takes the form
\begin{equation}\label{Second1M1}
\medmath{\left[\frac{d^{2}}{d\rho^{2}}+\frac{1+2\nu_{1}+2\nu_{2}}{\rho}\frac{d}{d\rho}-\rho^{2}-\frac{\lambda_-^{2}+4\nu_{1}\nu_{2}}{\rho^{2}}-2\lambda_-
+\frac{2B\mu_{B}g_{s}m_{s}}{\omega_{c}}\left(1+\nu_{1}-\nu_{2}\right)+\frac{4E}{\omega_{c}}\right]\phi_{\ell,m_s}^{+,-}(\rho)=0 },
\end{equation}
with
\begin{equation}\label{LAM2}
\lambda_{-} = \pm 2\sqrt{(\ell + \nu_{1})(\ell +\nu_{2})},\qquad \ell \in \{1/2,3/2,5/2,\ldots\}.
\end{equation}
In this sector, the angular quantum number $\ell$ takes half-integer values. This behavior originates from the asymmetric action of the reflection operators in the Dunkl angular equation, which modifies the admissible angular states and naturally leads to a semi-integer quantization of $\ell$, analogous to fermionic angular-momentum structures \cite{Genest2013,Genest2014}.

As a preliminary step toward the Schr\"odinger factorization procedure, we introduce the transformation defined in Eq.~(\ref{SecondC}) in order to eliminate the first-order derivative term appearing in Eq.~(\ref{Second1M1}), namely,
\begin{equation}
\left[\frac{d^{2}}{d\rho^{2}}-\rho^{2}+\frac{\lambda_-^{2}+(\nu_{1}-\nu_{2})^{2}-\frac{1}{4}}{\rho^{2}}-2\lambda_- +\frac{2B\mu_{B}g_{s}m_{s}}{\omega_{c}}\left(1+\nu_{1}-\nu_{2}\right)+\frac{4E}{\omega_{c}}
\right]\Psi_{\ell,m_s}^{+,-}(\rho)=0 .
\end{equation}

The quantization condition reads
\begin{equation}
-\lambda_{-}+\frac{B\mu_{B} g_{s} m_{s}}{\omega_{c}}\left(1+\nu_{1}-\nu_{2}\right)+\frac{2E}{\omega_{c}}=2(n_r+\ell)+(1+\nu_1+\nu_2),
\end{equation}

and the corresponding energy spectrum is
\begin{equation}
E^{+,-}_{n_r,\ell,m_s}=\omega_c\left[n_r+\ell+\frac{1+\nu_1+\nu_2}{2}+\sqrt{(\ell+\nu_1)(\ell+\nu_2)}-\frac{m_s(1+\nu_1-\nu_2)}{2}\right].
\end{equation}
Physically, this sector highlights how the asymmetry between the deformation parameters modifies the effective interaction, leading to distinct spectral shifts compared with the symmetric sectors.

\subsection{Case $\epsilon=-1$ $(\epsilon_1=-1,\epsilon_2=1)$}
This sector is complementary to the previous one, with the roles of $\nu_1$ and $\nu_2$ interchanged. The radial equation is
\begin{equation}
\medmath{\left[\frac{d^{2}}{d\rho^{2}}+\frac{1+2\nu_{1}+2\nu_{2}}{\rho}\frac{d}{d\rho}-\rho^{2}-\frac{\lambda_-^{2}-4\nu_1\nu_2}{\rho^{2}}-2\lambda_-+\frac{2B\mu_B g_s m_s}{\omega_c}\left(1+\nu_2-\nu_1\right)
+\frac{4E}{\omega_c}\right]\phi_{\ell,m_s}^{-,+}(\rho)=0}.
\end{equation}

After transformation,
\begin{equation}
\left[\frac{d^{2}}{d\rho^{2}}-\rho^{2}+\frac{\lambda_-^{2}+(\nu_{1}-\nu_{2})^{2}-\frac14}{\rho^{2}}-2\lambda_-+\frac{2B\mu_B g_s m_s}{\omega_c}\left(1+\nu_2-\nu_1\right)
+\frac{4E}{\omega_c}\right]\Psi_{\ell,m_s}^{-,+}(\rho)=0.
\end{equation}
The quantization condition becomes
\begin{equation}
-\lambda_{-}+\frac{B\mu_{B} g_{s} m_{s}}{\omega_{c}}\left(1+\nu_{2}-\nu_{1}\right)+\frac{2E}{\omega_{c}}=2(n_r+\ell)+(1+\nu_2-\nu_1),
\end{equation}
leading to the energy spectrum
\begin{equation}
E^{-,+}_{n_r,\ell,m_s}=\omega_c\left[n_r+\ell+\frac{1+\nu_1+\nu_2}{2}+\sqrt{(\ell+\nu_1)(\ell+\nu_2)}-\frac{m_s(1+\nu_2-\nu_1)}{2}\right].
\end{equation}
It is worth emphasizing that, in the undeformed limit $\nu_1=\nu_2=0$, all reflection sectors $(\epsilon_1,\epsilon_2)$ recover the same energy spectrum given in Eq.~(\ref{ENERNUC}).

This result confirms that the interchange $\nu_1 \leftrightarrow \nu_2$ produces a corresponding interchange in the spectral corrections, while preserving the same algebraic structure.
\begin{figure}[H]
    \centering
    \includegraphics[width=0.97\textwidth]{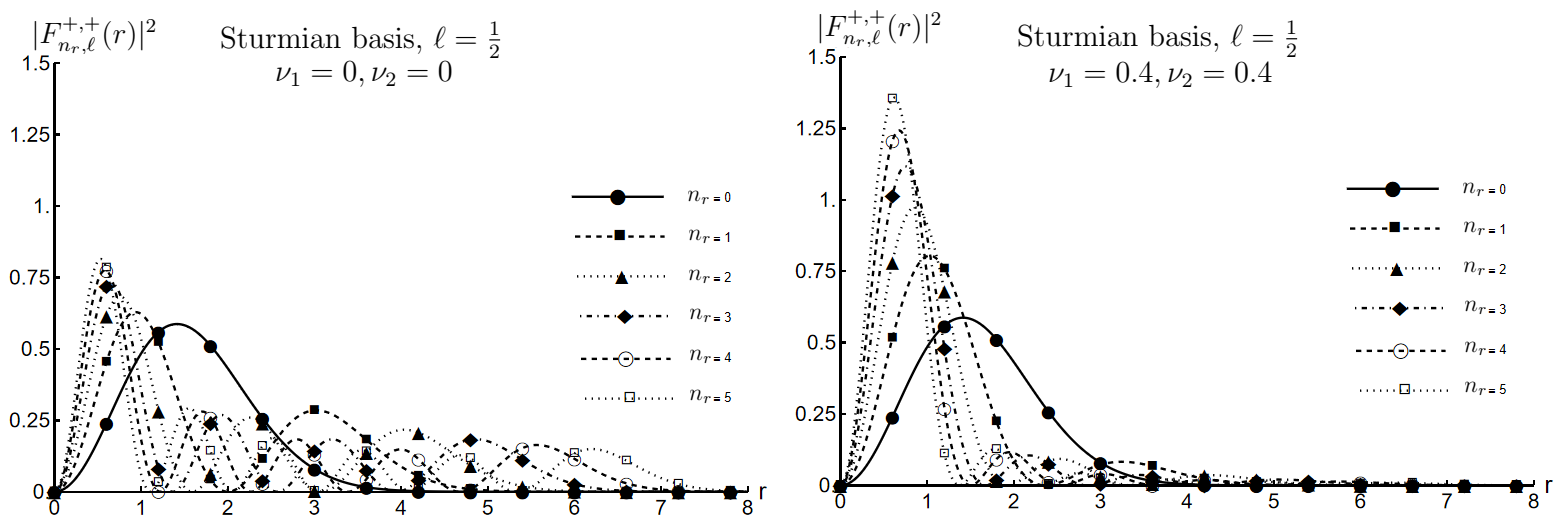}
    \caption{Normalized radial probability density $|F_{n_r,\ell}^{+,+}(r)|^2$
of the Sturmian basis for $\ell=\frac{1}{2}$, $(\nu_1,\nu_2)=(0,0)$ and $(0.4,0.4)$,
with $n_r=0,1,\ldots,5$.}
\end{figure}
\begin{figure}[H]
    \centering
    \includegraphics[width=0.97\textwidth]{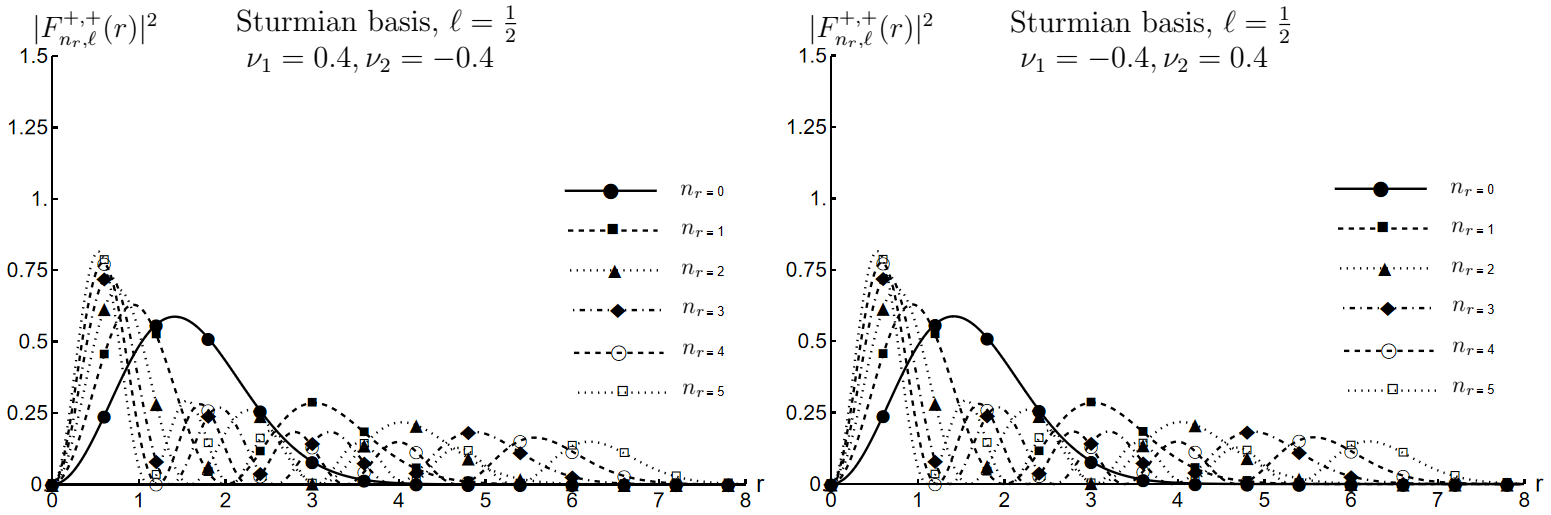}
    \caption{Normalized radial probability density $|F_{n_r,\ell}^{+,+}(r)|^2$
of the Sturmian basis for $\ell=\frac{1}{2}$, $(\nu_1,\nu_2)=(0.4,-0.4)$ and $(-0.4,0.4)$,
with $n_r=0,1,\ldots,5$.}
\end{figure}

\begin{figure}[H]
    \centering
    \includegraphics[width=0.50\textwidth]{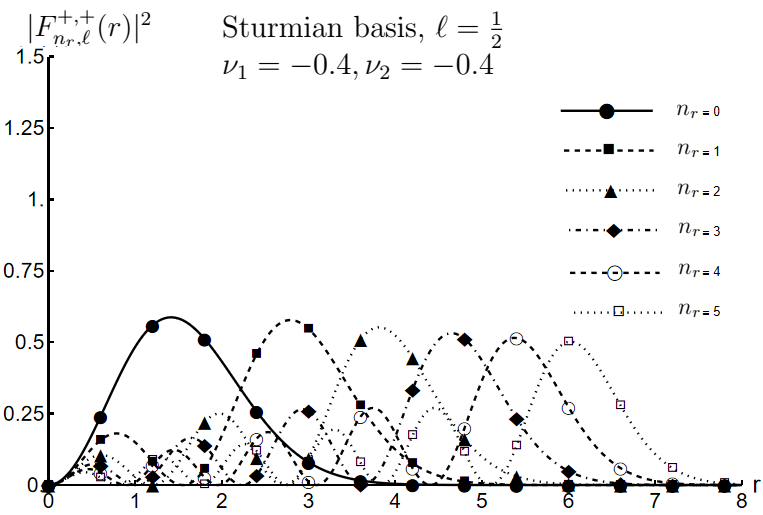}
    \caption{Normalized radial probability density $|F_{n_r,\ell}^{+,+}(r)|^2$
of the Sturmian basis for $\ell=\frac{1}{2}$, $(\nu_1,\nu_2)=(-0.4,-0.4)$ , with $n_r=0,1,\ldots,5$.}
\end{figure}

Figs.~$12$--$14$ display the normalized radial probability density $|F_{n_r,\ell}^{+,+}(r)|^2$ for $\ell=\tfrac12$ and
$n_r=0,\dots,5$ in the same deformation sectors considered in Figs.~$1$--$3$. The qualitative behavior with respect to $n_r$ remains the same:
the number of nodes increases and the principal maximum moves toward larger $r$ as the radial excitation grows. However, in comparison with the
$\ell=1$ case, all distributions are shifted closer to the origin, i.e., toward smaller radial distances.

This difference arises from the reduction of the effective centrifugal barrier, since the Bargmann index
$k=\ell+\tfrac{\nu_1+\nu_2+1}{2}$ decreases when $\ell=\tfrac12$. Consequently, the particle experiences weaker repulsion near $r=0$ and the
probability density becomes more localized at small radii, while the dependence on the deformation parameter $\nu_1+\nu_2$ preserves the same
trend observed in Figs.~$1$--$3$. This confirms the role of $\ell$ as an effective angular momentum governing the radial Dunkl dynamics.
\begin{figure}[H]
    \centering
    \includegraphics[width=0.5\textwidth]{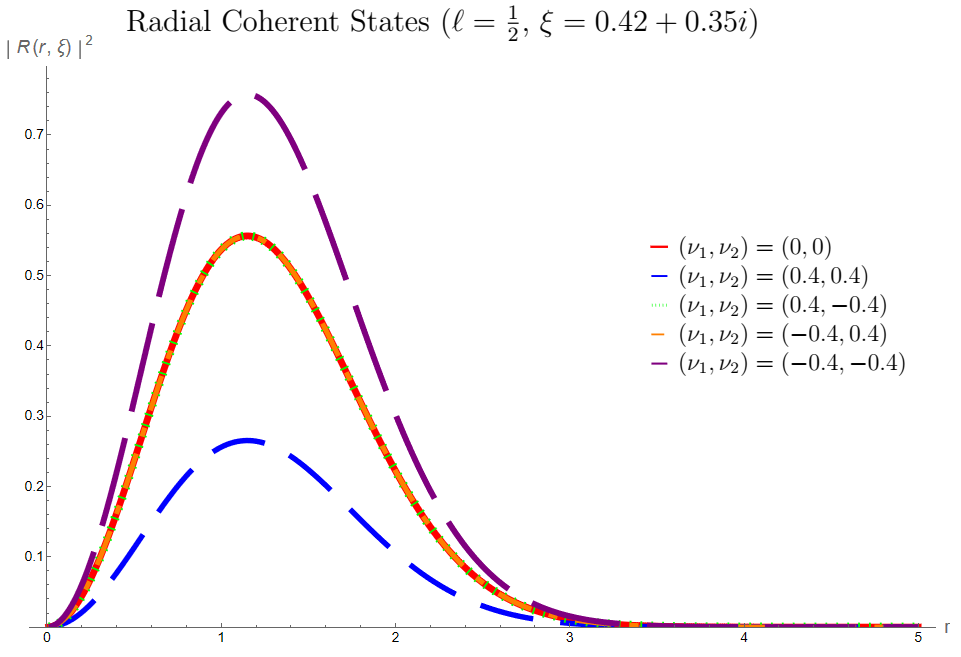}
    \caption{Radial coherent-state probability density $|R(r,\xi)|^2$ for $\ell=\frac{1}{2}$,
$\xi=0.42+0.35i$, and $(\nu_1,\nu_2)=(0,0),(0.4,0.4),(0.4,-0.4),(-0.4,0.4),(-0.4,-0.4)$.}
\end{figure}

\begin{figure}[H]
    \centering
    \includegraphics[width=0.50\textwidth]{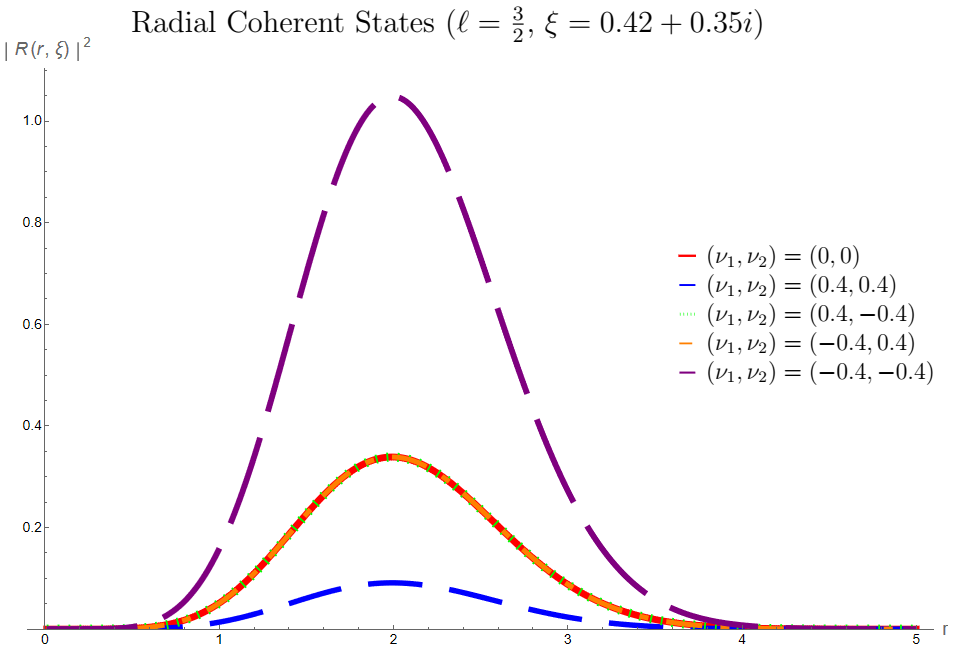}
    \caption{Radial coherent-state probability density $|R(r,\xi)|^2$ for $\ell=\frac{3}{2}$,
$\xi=0.42+0.35i$, and $(\nu_1,\nu_2)=(0,0),(0.4,0.4),(0.4,-0.4),(-0.4,0.4),(-0.4,-0.4)$.}
\end{figure}

\begin{figure}[H]
    \centering
    \includegraphics[width=0.50\textwidth]{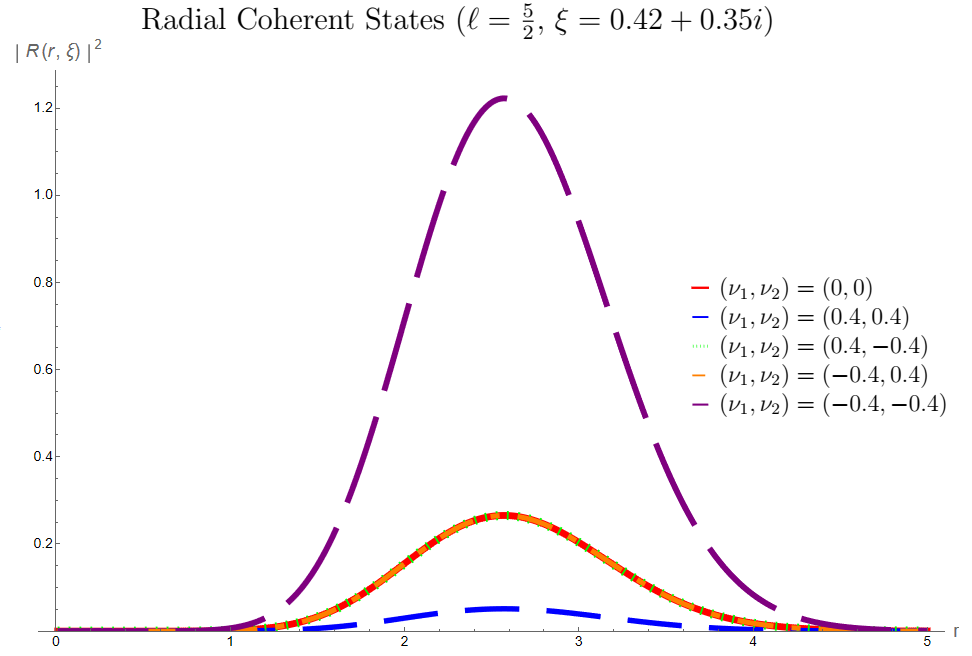}
    \caption{Radial coherent-state probability density $|R(r,\xi)|^2$ for $\ell=\frac{5}{2}$,
$\xi=0.42+0.35i$, and $(\nu_1,\nu_2)=(0,0),(0.4,0.4),(0.4,-0.4),(-0.4,0.4),(-0.4,-0.4)$.}
\end{figure}

\begin{figure}[H]
    \centering
    \includegraphics[width=0.50\textwidth]{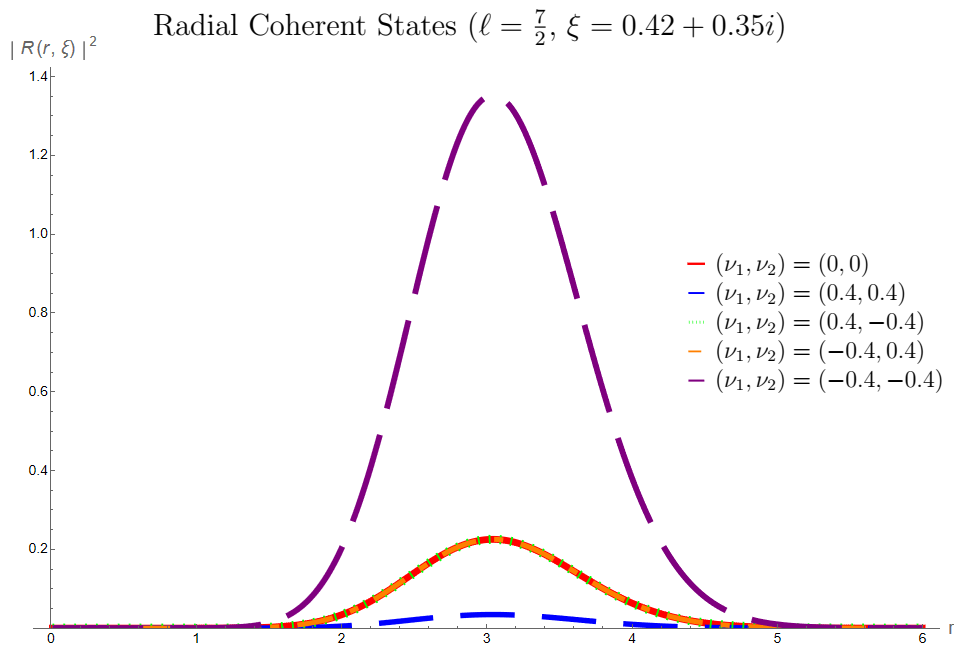}
    \caption{Radial coherent-state probability density $|R(r,\xi)|^2$ for $\ell=\frac{7}{2}$,
$\xi=0.42+0.35i$, and $(\nu_1,\nu_2)=(0,0),(0.4,0.4),(0.4,-0.4),(-0.4,0.4),(-0.4,-0.4)$.}
\end{figure}

\begin{figure}[H]
    \centering
    \includegraphics[width=0.50\textwidth]{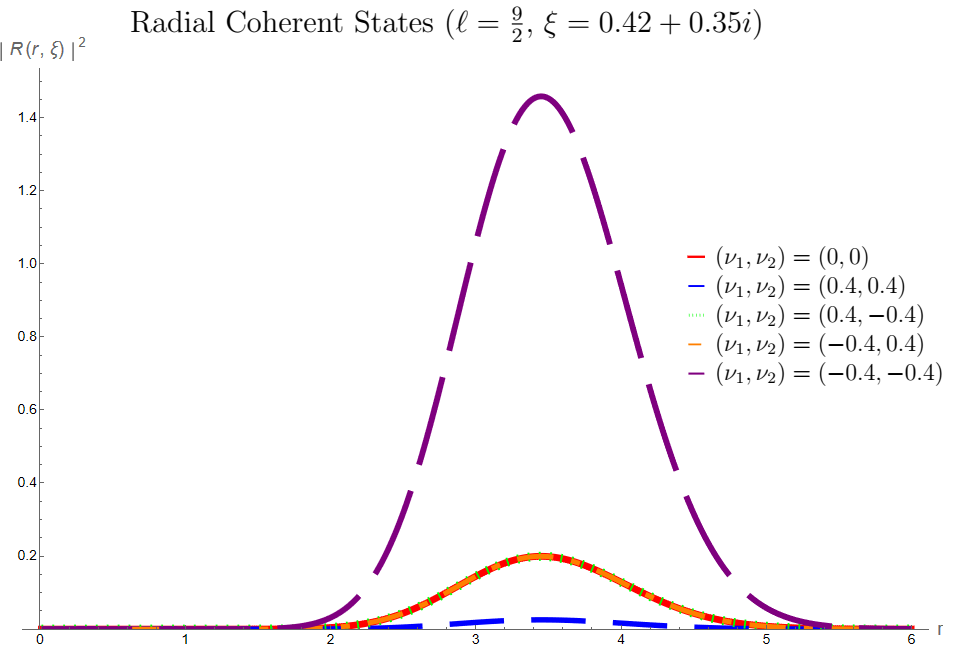}
    \caption{Radial coherent-state probability density $|R(r,\xi)|^2$ for $\ell=\frac{9}{2}$,
$\xi=0.42+0.35i$, and $(\nu_1,\nu_2)=(0,0),(0.4,0.4),(0.4,-0.4),(-0.4,0.4),(-0.4,-0.4)$.}
\end{figure}
Figs.~$15$--$19$ display the radial coherent-state density $|R(r,\xi)|^{2}$ for half-integer orbital quantum numbers
$\ell=\tfrac12,\tfrac32,\tfrac52,\tfrac72,\tfrac92$. As $\ell$ increases, the maximum of the distribution progressively shifts
toward larger radii and develops a ring-like profile, since the factor $r^{4\ell}$ suppresses the probability near the origin while the exponential
term preserves the overall confinement. Moreover, some curves overlap because they depend only on the effective combination $\nu_1+\nu_2$.
In contrast with Figs.~$4$--$8$, where integer values of $\ell$ produce more noticeable radial changes between consecutive panels, the present case shows a
smoother and more continuous variation of the mean radius, highlighting the quasiclassical role of $\ell$ as an effective angular momentum controlling
the orbital size of the coherent packet.
\begin{figure}[H]
    \centering
    \includegraphics[width=0.92\textwidth]{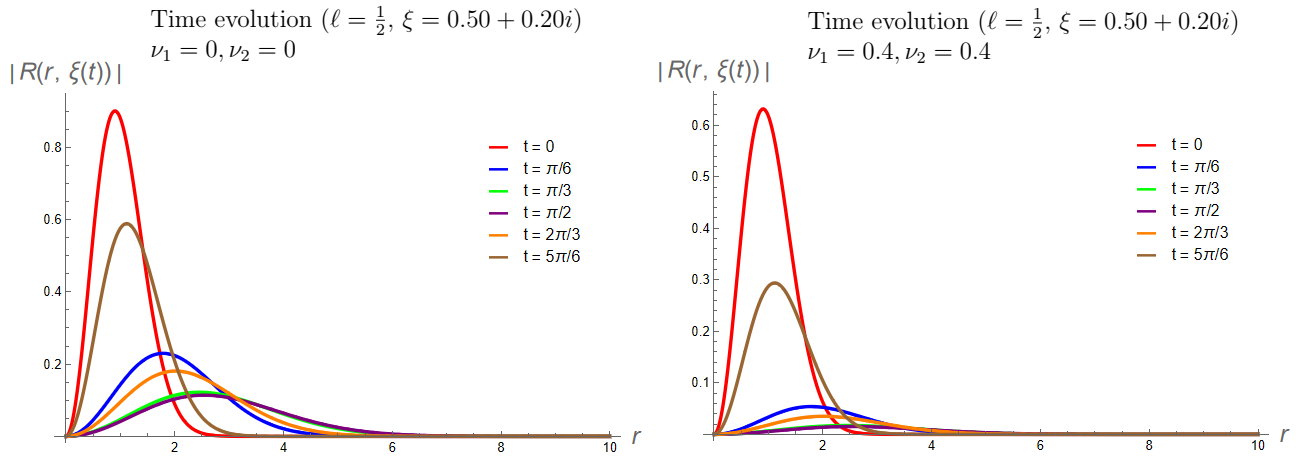}
    \caption{Time evolution of the radial coherent-state probability density $|R(r,\xi(t))|^{2}$ for $\ell=\frac{1}{2}$, $\xi=0.50+0.20i$, and $(\nu_1,\nu_2)=(0,0)$ and $(0.4,0.4)$ at $t=0,\pi/6,\pi/3,\pi/2,2\pi/3,5\pi/6$.}
\end{figure}

\begin{figure}[H]
    \centering
    \includegraphics[width=1.00\textwidth]{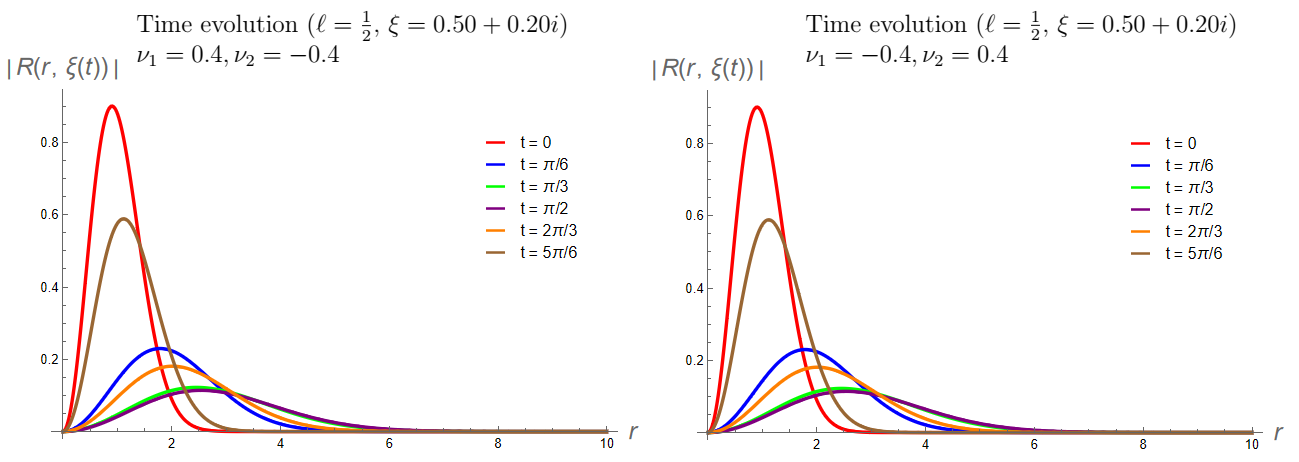}
    \caption{Time evolution of the radial coherent-state probability density $|R(r,\xi(t))|^{2}$ for $\ell=\frac{1}{2}$, $\xi=0.50+0.20i$, and $(\nu_1,\nu_2)=(0.4,-0.4)$ and $(-0.4,0.4)$ at $t=0,\pi/6,\pi/3,\pi/2,2\pi/3,5\pi/6$.}
\end{figure}

\begin{figure}[H]
    \centering
    \includegraphics[width=0.54\textwidth]{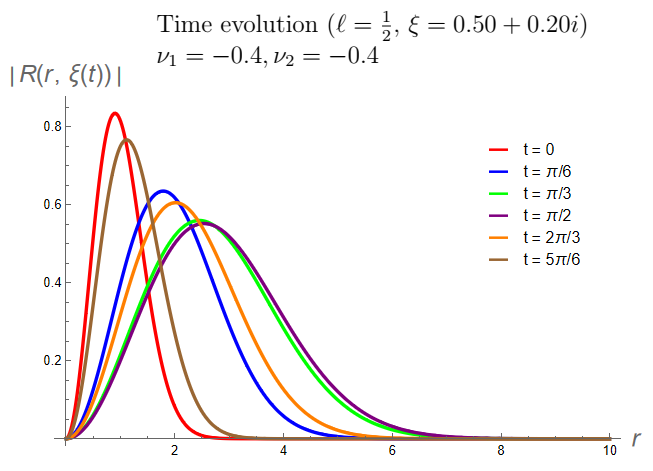}
    \caption{Time evolution of the radial coherent-state probability density $|R(r,\xi(t))|^{2}$ for $\ell=\frac{1}{2}$, $\xi=0.50+0.20i$, and $(\nu_1,\nu_2)=(-0.4,-0.4)$ at $t=0,\pi/6,\pi/3,\pi/2,2\pi/3,5\pi/6$.}
\end{figure}
Figs.~$20$--$22$ present the time evolution of $|R(r,\xi(t))|^{2}$ for $\ell=\tfrac12$, in contrast with Figs.~$9$--$11$ where $\ell=1$.
In both cases, the same radial ``breathing'' mode is observed, characterized by a periodic oscillation of the profile in which the probability maximum
smoothly shifts within a temporal cycle. However, when passing from $\ell=1$ to $\ell=\tfrac12$, the entire dynamics occurs at smaller radial
distances due to the weakening of the effective centrifugal barrier associated with the orbital angular momentum. Although the periodicity
and qualitative structure of the evolution remain unchanged, the mean radius around which the coherent packet oscillates depends explicitly on
both $\ell$ and the effective combination $\nu_1+\nu_2$. This comparison confirms that $\ell$ plays the role of an effective angular momentum
controlling the orbital size of the coherent state without modifying the intrinsic nature of its temporal dynamics.

It is worth emphasizing that all results presented in Secs.~3 and 4 assume $|\nu_1| = |\nu_2|$, which guarantees that the radial dynamics depends exclusively on the sum $\nu_1 + \nu_2$. This property, which leads to the degeneracy observed in the figures, will be broken in the asymmetric case $|\nu_1| \neq |\nu_2|$ analyzed in Sec.~5.

\section{Graphical Representation for the Asymmetric Case $(|\nu_1| \neq |\nu_2|)$}
\begin{figure}[H]
    \centering
    \includegraphics[width=0.91\textwidth]{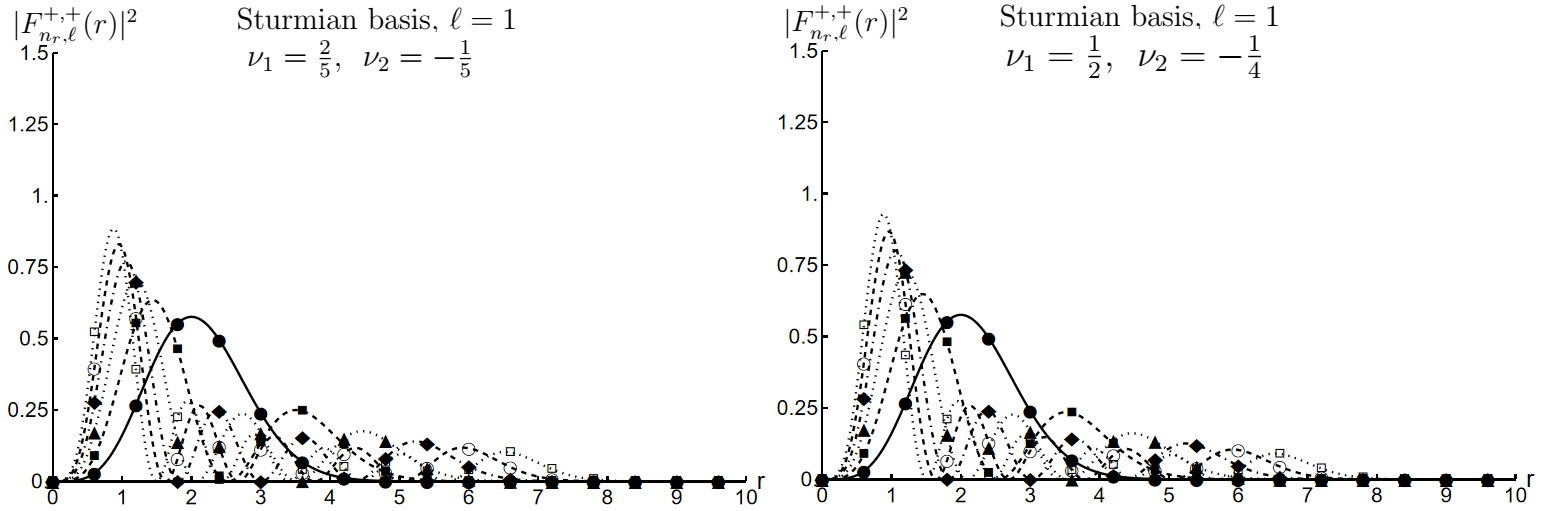}
    \caption{Normalized radial probability density $|F_{n_r,\ell}^{+,+}(r)|^2$
of the Sturmian basis for $\ell=1$, $(\nu_1,\nu_2)=(\frac{2}{5},-\frac{1}{5})$ and $(\frac{1}{2},-\frac{1}{4})$,
with $n_r=0,1,\ldots,5$.}
\end{figure}
For simplicity and compactness, only a few representative examples corresponding to the asymmetric Dunkl parameter configurations ($|\nu_1|\neq|\nu_2|$) are presented, as illustrated in Fig.~$23$. The radial probability densities preserve the same qualitative behavior observed in the symmetric configurations: as the radial quantum number $n_r$ increases, the principal maximum shifts toward larger values of $r$ and the number of nodes grows progressively.

The asymmetric deformations introduce only mild modifications in the amplitude and localization of the curves, without altering the overall structure of the profiles, mainly producing a gradual adjustment of the radial distribution. In contrast with the symmetric case $(|\nu_1|=|\nu_2|)$, where all radial expressions depend solely on $\nu_1+\nu_2$, here the individual values of $\nu_1$ and $\nu_2$ become relevant, breaking the degeneracy observed in previous figures.

\begin{figure}[H]
    \centering
    \includegraphics[width=0.91\textwidth]{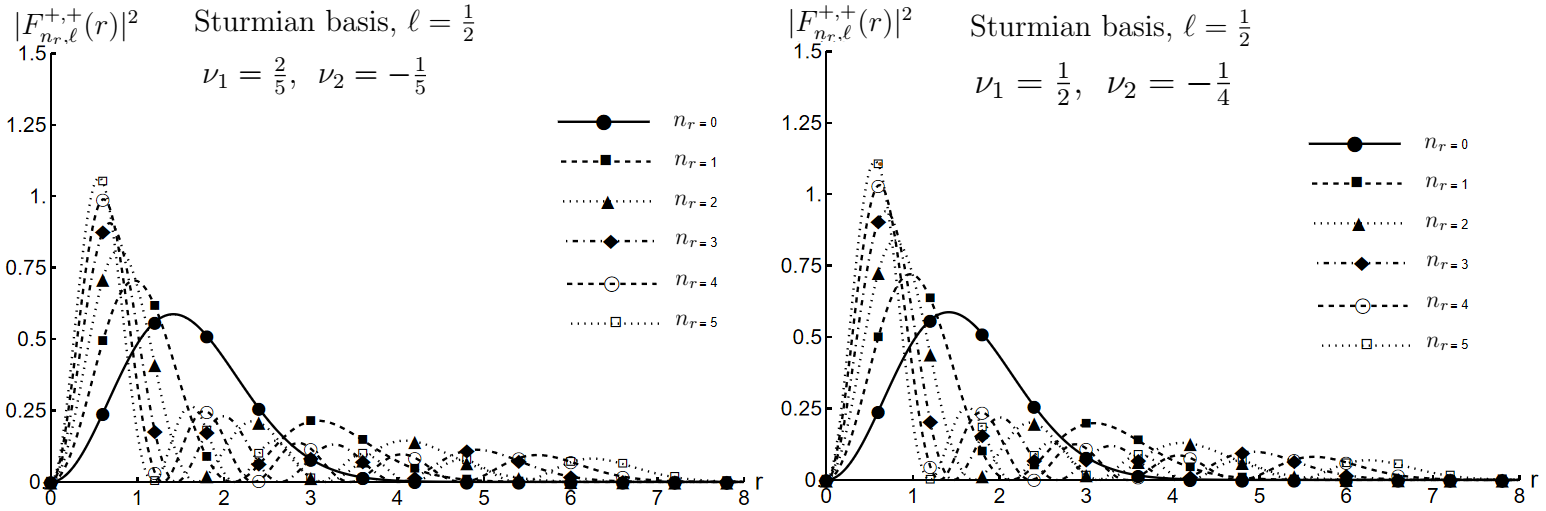}
    \caption{Normalized radial probability density $|F_{n_r,\ell}^{+,+}(r)|^2$
of the Sturmian basis for $\ell=\frac{1}{2}$, $(\nu_1,\nu_2)=(\frac{2}{5},-\frac{1}{5})$ and $(\frac{1}{2},-\frac{1}{4})$,
with $n_r=0,1,\ldots,5$.}
\end{figure}
Several asymmetric configurations of the Dunkl parameters were analyzed. However, in order to avoid unnecessarily extending the present work, only two representative cases are included here, as shown in Fig.~$24$, since they adequately illustrate the main effects of the asymmetric Dunkl deformations on the radial probability distributions. The asymmetric configurations of the Dunkl parameters ($|\nu_1|\neq|\nu_2|$) for $\ell=\tfrac{1}{2}$ preserve the same qualitative behavior observed in the symmetric case. As the radial quantum number $n_r$ increases, the principal maximum of the radial density shifts toward larger values of $r$ and the number of nodes grows progressively. The radial localization is mainly governed by the combination $\nu_1+\nu_2$, while the asymmetry between $\nu_1$ and $\nu_2$ introduces only mild modifications in the amplitude and shape of the profiles, without altering the overall structure of the radial distribution.
\begin{figure}[H]
    \centering
    \includegraphics[width=0.5\textwidth]{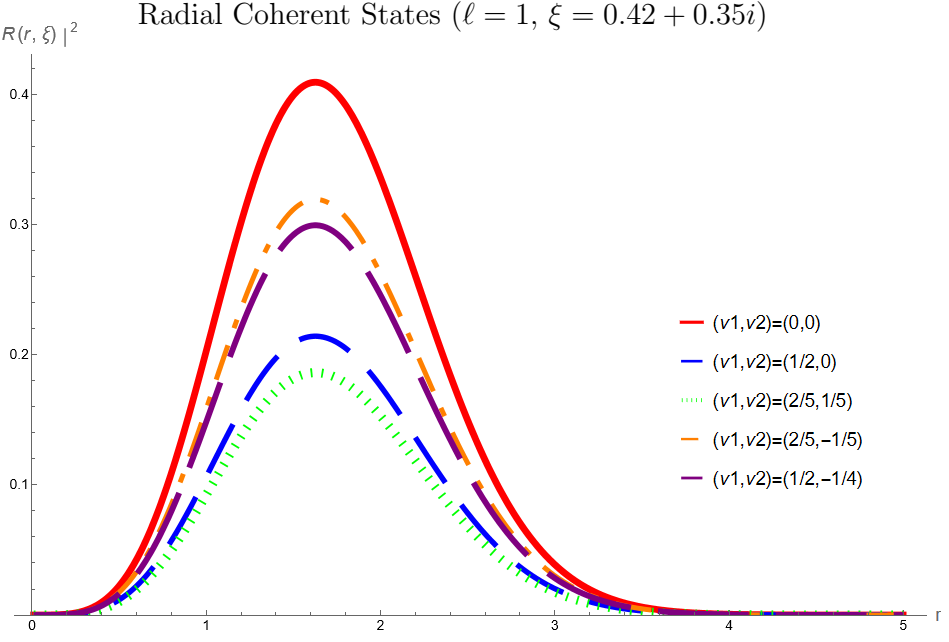}
    \caption{Radial coherent-state probability density $|R(r,\xi)|^2$ for $\ell=1$,
$\xi=0.42+0.35i$, and $(\nu_1,\nu_2)=(0,0),(\frac{1}{2},0),(\frac{2}{5},\frac{1}{5}),(\frac{2}{5},-\frac{1}{5}),(\frac{1}{2},-\frac{1}{4})$.}
\end{figure}

\begin{figure}[H]
    \centering
    \includegraphics[width=0.50\textwidth]{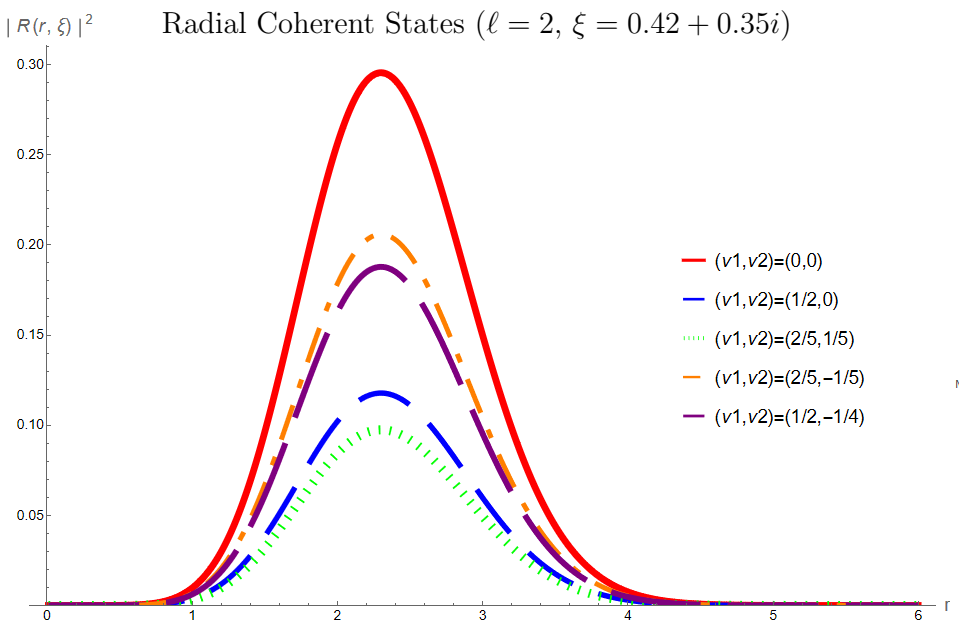}
    \caption{Radial coherent-state probability density $|R(r,\xi)|^2$ for $\ell=2$,
$\xi=0.42+0.35i$, and $(\nu_1,\nu_2)=(0,0),(\frac{1}{2},0),(\frac{2}{5},\frac{1}{5}),(\frac{2}{5},-\frac{1}{5}),(\frac{1}{2},-\frac{1}{4})$.}
\end{figure}
Although several values of $\ell$ were analyzed, only the representative cases $\ell=1$ and $\ell=2$ are included here, as illustrated in Figs.~$25$ and $26$, in order to avoid making the manuscript unnecessarily more extensive. For these cases, the asymmetric configurations of the Dunkl parameters preserve the same qualitative behavior observed in the symmetric case: as $\ell$ increases, the radial distributions shift toward larger values of $r$ and develop a more pronounced ring-like profile due to the increasing effective centrifugal barrier. However, unlike the symmetric case, where some curves may overlap, in the asymmetric configurations each pair $(\nu_1,\nu_2)$ generates a distinct radial profile, showing that the asymmetry removes the degeneracy present in the symmetric configurations.
\begin{figure}[H]
    \centering
    \includegraphics[width=0.96\textwidth]{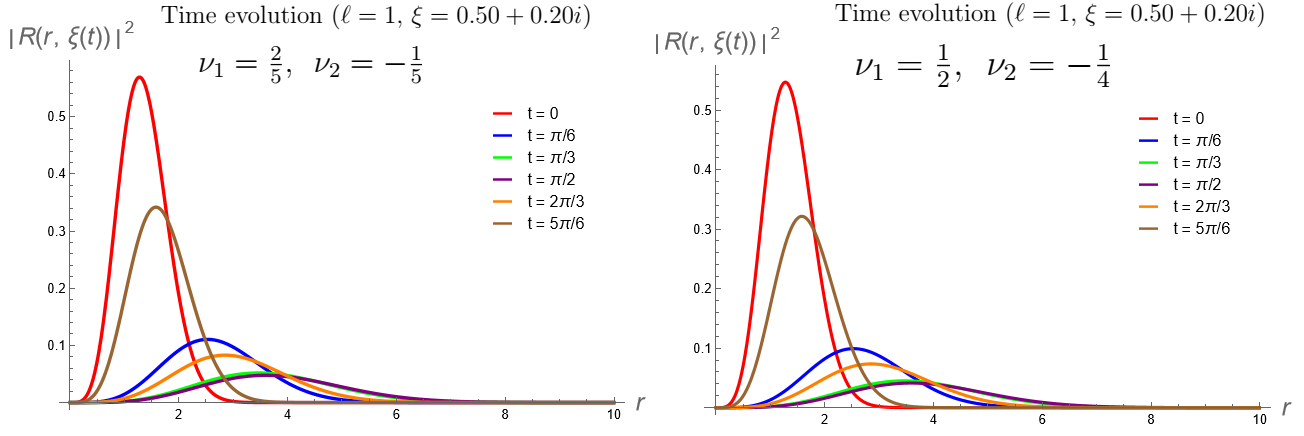}
    \caption{Time evolution of the radial coherent-state probability density $|R(r,\xi(t))|^{2}$ for $\ell=1$, $\xi=0.50+0.20i$, and $(\nu_1,\nu_2)=(\frac{2}{5},-\frac{1}{5})$ and $(\frac{1}{2},-\frac{1}{4})$ at $t=0,\pi/6,\pi/3,\pi/2,2\pi/3,5\pi/6$.}
\end{figure}
The figures corresponding to the asymmetric case, shown in Fig.~$27$, display the time evolution of $|R(r,\xi(t))|^{2}$ for $\ell=1$. As in the symmetric case, a periodic oscillation of the radial profile is observed, in which the probability maximum smoothly shifts along the radial coordinate during the temporal cycle. The global structure and periodicity of the dynamics remain unchanged, confirming the robustness of the coherent behavior.

The differences between symmetric and asymmetric configurations manifest slightly in the amplitude and in the width of the distributions, which depend explicitly on the parameters $(\nu_1,\nu_2)$.
\begin{figure}[H]
    \centering
    \includegraphics[width=0.5\textwidth]{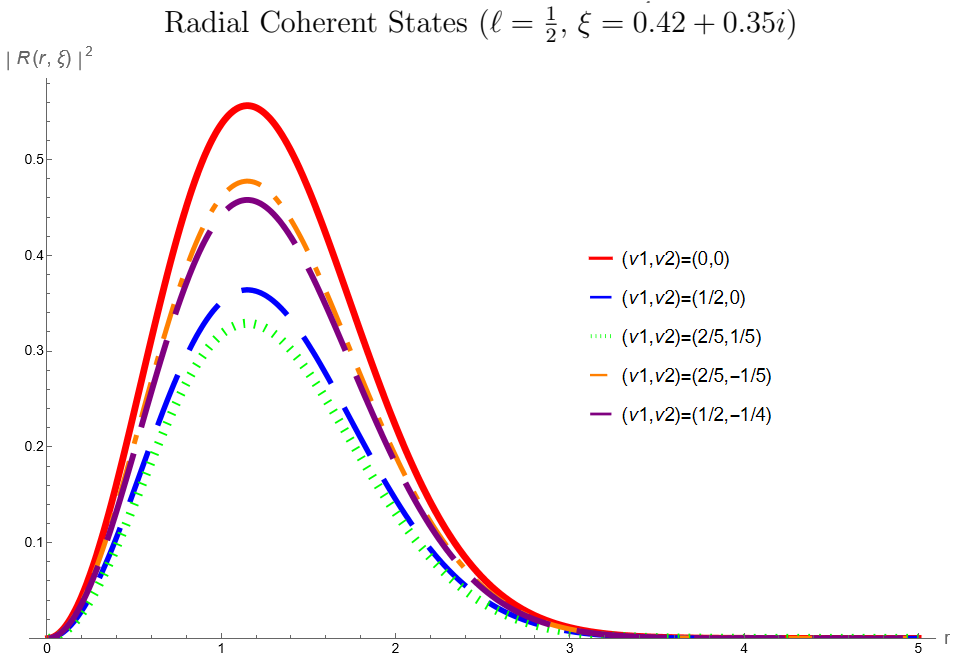}
    \caption{Radial coherent-state probability density $|R(r,\xi)|^2$ for $\ell=1$,
$\xi=0.42+0.35i$, and $(\nu_1,\nu_2)=(0,0),(\frac{1}{2},0),(\frac{2}{5},\frac{1}{5}),(\frac{2}{5},-\frac{1}{5}),(\frac{1}{2},-\frac{1}{4})$.}
\end{figure}

\begin{figure}[H]
    \centering
    \includegraphics[width=0.50\textwidth]{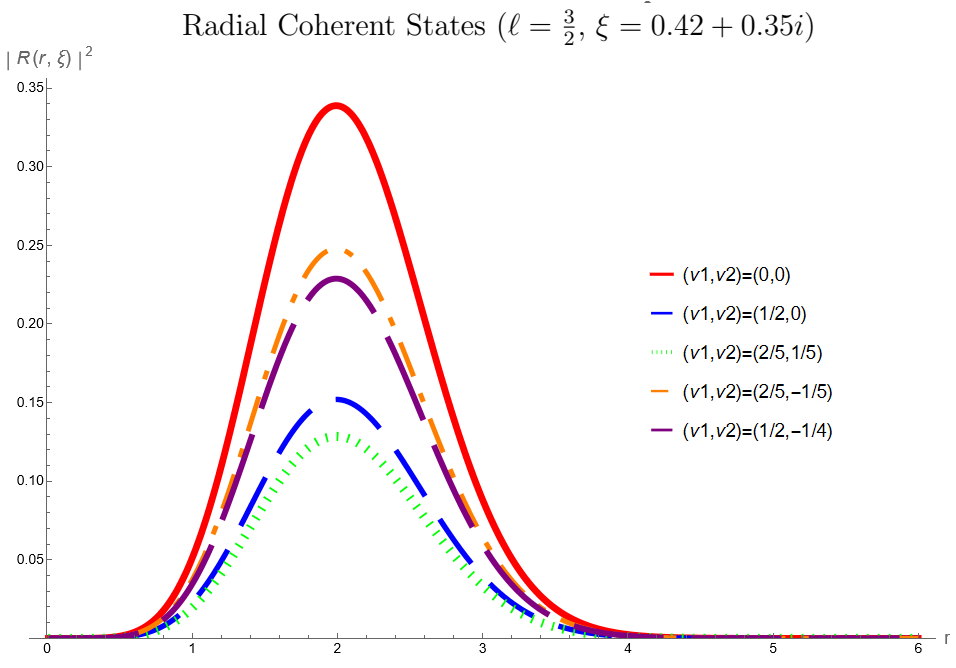}
    \caption{Radial coherent-state probability density $|R(r,\xi)|^2$ for $\ell=2$,
$\xi=0.42+0.35i$, and $(\nu_1,\nu_2)=(0,0),(\frac{1}{2},0),(\frac{2}{5},\frac{1}{5}),(\frac{2}{5},-\frac{1}{5}),(\frac{1}{2},-\frac{1}{4})$.}
\end{figure}
Although several half-integer values of $\ell$ were analyzed, only the representative cases $\ell=\tfrac12$ and $\ell=\tfrac32$ are presented here, as illustrated in Figs.~28 and 29, in order to keep the discussion compact. The radial coherent states for half-integer orbital quantum numbers and asymmetric Dunkl parameter configurations preserve the same qualitative behavior observed in the symmetric case: as $\ell$ increases, the distributions shift toward larger values of $r$ due to the growth of the effective centrifugal barrier. However, unlike the symmetric configurations, where several curves overlap because of their dependence on the combination $\nu_1+\nu_2$, in the asymmetric case each pair $(\nu_1,\nu_2)$ generates a distinct radial profile. This shows that the asymmetry removes the degeneracy present in the symmetric case while preserving the overall radial dynamics.
\begin{figure}[H]
    \centering
    \includegraphics[width=0.96\textwidth]{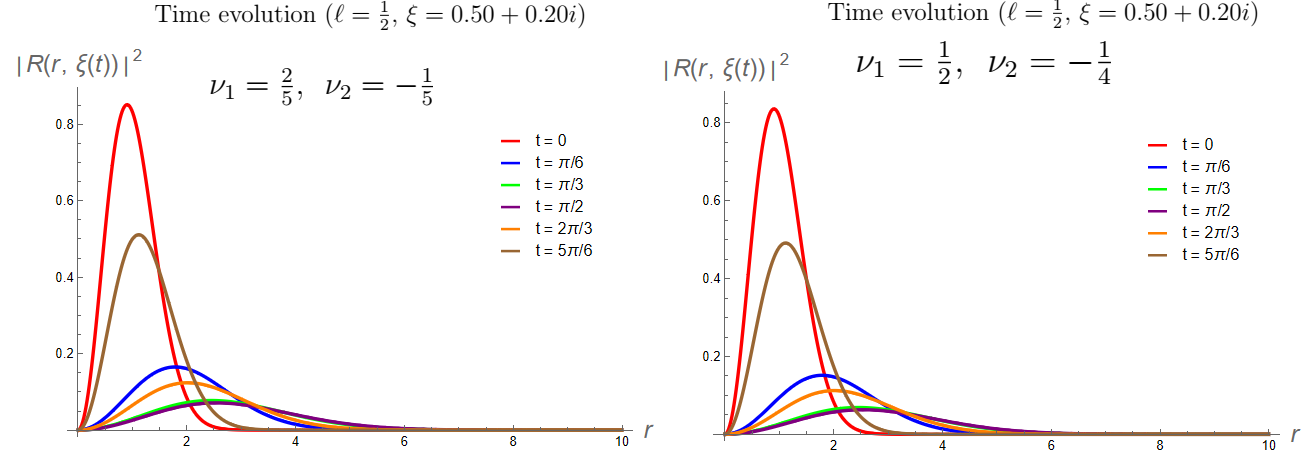}
    \caption{Time evolution of the radial coherent-state probability density $|R(r,\xi(t))|^{2}$ for $\ell=\frac{1}{2}$, $\xi=0.50+0.20i$, and $(\nu_1,\nu_2)=(\frac{2}{5},-\frac{1}{5})$ and $(\frac{1}{2},-\frac{1}{4})$ at $t=0,\pi/6,\pi/3,\pi/2,2\pi/3,5\pi/6$.}
\end{figure}
The time evolution of $|R(r,\xi(t))|^{2}$ for fractional Dunkl parameters with $\ell=\tfrac{1}{2}$, shown in Fig.~30, exhibits the same qualitative behavior as in the $\ell=1$ case: the radial profile evolves smoothly in time, with the probability maximum shifting within a finite radial region while preserving its shape and periodicity.

The main difference arises from the weaker effective centrifugal barrier for $\ell=\tfrac{1}{2}$, which shifts the dynamics toward smaller radial distances and leads to higher peaks due to stronger radial localization. In both cases, fractional Dunkl parameters introduce only slight variations in amplitude and width, without altering the overall structure of the dynamics.

In contrast with the symmetric case ($|\nu_1| = |\nu_2|$), where all radial quantities depend solely on $\nu_1 + \nu_2$, in the asymmetric case the individual values of $\nu_1$ and $\nu_2$ become relevant. Consequently, pairs with the same sum but different individual parameters yield distinct radial profiles, breaking the degeneracy observed in previous sections.

For $\epsilon=+1$ (i.e., $\epsilon_{1}=\epsilon_{2}=1$ or $\epsilon_{1}=\epsilon_{2}=-1$) and $\epsilon=-1$
(i.e., $\epsilon_{1}=1,\epsilon_{2}=-1$ or $\epsilon_{1}=-1,\epsilon_{2}=1$), the $\mathfrak{su}(1,1)$
algebraic framework yields identical radial Sturmian bases, since all sectors share the same quadratic Casimir invariant
and Bargmann index $k$ (see Eq.~(\ref{CASC1})). Consequently, the coherent states and their time evolution (see Eqs.~(\ref{COHE1}) and (\ref{EVTEC}))
remain formally unchanged, with differences appearing only in the corresponding energy spectra.

Quantitative distinctions arise, however, in the numerical and graphical analysis due to the different allowed values of the
orbital quantum number $\ell$ in each sector (see Eqs.~(\ref{LAM1}) and (\ref{LAM2})). Although the $\mathfrak{su}(1,1)$ representation is formally the same,
the integer versus half-integer character of $\ell$ modifies the effective centrifugal barrier and thus the radial localization scale.
This accounts for the different radial ranges observed in the figures while preserving the same underlying dynamical structure.

As a consistency check of the present formalism, we consider the non-deformed limit corresponding to $\nu_1=\nu_2=0$. In this regime, the expressions obtained for the radial wave functions, the energy spectrum, the Sturmian basis, and the coherent states together with their time evolution reduce exactly to the well-known results of the ordinary Pauli oscillator in the presence of a uniform magnetic field. This limiting behavior shows that the Dunkl parameters introduce a continuous deformation of the standard system while preserving the underlying $\mathrm{SU}(1,1)$ algebraic structure, in agreement with results reported in the literature \cite{Perelomov,Barut1,RosasOrtiz,OjedaGuillen}.

\section{Concluding Remarks}
In this work, we have developed an algebraic solution of the two-dimensional Dunkl--Pauli equation for a spin--$1/2$ particle in the presence of a uniform magnetic field. The inclusion of Dunkl derivatives leads to reflection-dependent sectors and induces parity-sensitive quantum dynamics. We have shown that, for each reflection sector, the radial part of the Dunkl--Pauli equation admits an underlying $\mathfrak{su}(1,1)$ Lie algebraic structure.

By applying the Schr\"odinger factorization method, the generators of the $\mathfrak{su}(1,1)$ algebra were constructed explicitly, and the energy spectra were obtained through the theory of irreducible unitary representations. Although the energy spectra depend on the reflection parameters, all sectors are characterized by the same quadratic Casimir invariant and share an identical Bargmann index, which gives rise to a common Sturmian radial basis. Furthermore, this algebraic framework allows for the construction of $\mathrm{SU}(1,1)$ Perelomov radial coherent states and the analysis of their time evolution. These results highlight the unifying role of $\mathfrak{su}(1,1)$ symmetry in the algebraic description of the Dunkl--Pauli equation in a magnetic field.

The graphical analysis further elucidates the role of the Dunkl deformation and the orbital quantum number $\ell$ in both the stationary and
time-dependent radial dynamics. The Sturmian densities show that the combined effect of $\ell$ and $\nu_1+\nu_2$ directly controls the spatial
localization of the states, shifting the probability distribution toward larger or smaller radii depending on the sign of $\nu_1+\nu_2$. In the
coherent sector, the evolution exhibits a periodic radial ``breathing'' mode whose mean radius depends on these same parameters, while the
temporal frequency remains determined solely by the phase rotation of the coherent parameter $\xi(t)$. Altogether, these results demonstrate that
the Dunkl deformation modifies the spatial structure of the states without altering the intrinsic periodic character of the dynamics,
reinforcing the organizing role of the underlying $\mathfrak{su}(1,1)$ symmetry in the Dunkl--Pauli system.

\noindent It is important to mention that an extension to non-reducible reflection groups $W$ may affect the separability of the system, since such groups introduce couplings between spatial coordinates. In this case, it may become more difficult to reduce the problem to a simple radial equation as in the present formulation. Furthermore, the structure of the underlying symmetry algebra could also be modified, in the sense that the $\mathfrak{su}(1,1)$ symmetry may not be realized in the same form. These effects could, in turn, impact the construction of the Sturmian basis, coherent states, and their time evolution. However, a detailed assessment of these effects lies beyond the scope of the present work and is left for future investigation.

The extension of the Dunkl--Pauli equation in the presence of a magnetic field to higher dimensions appears feasible within the Dunkl framework, which is well established in arbitrary dimensions for scalar systems. However, the inclusion of spin degrees of freedom and the tensorial nature of the magnetic field introduce additional complexities that may affect separability and the underlying algebraic structure. Consequently, such a generalization requires a careful formulation and is left for future investigation.

\section*{Appendix. Basic aspects of the $\mathrm{SU}(1,1)$ algebra, coherent states, and their time evolution}

\renewcommand{\theequation}{A\arabic{equation}}
\setcounter{equation}{0}

Lie algebras provide a powerful algebraic framework for the study of continuous symmetries in quantum systems. In this approach, the spectral properties of a physical model can be obtained through commutation relations and representation theory instead of solving the associated differential equations directly. In particular, the non-compact algebra $\mathfrak{su}(1,1)$ appears in a broad class of physical problems, ranging from relativistic quantum mechanics to quantum optics and squeezed states \cite{Yesiltas,Gazeau}. The generators of the $\mathfrak{su}(1,1)$ algebra satisfy the commutation relations \cite{Barut1}
\begin{equation}
[\mathcal{J}_1,\mathcal{J}_2]=-i\mathcal{J}_3,\qquad
[\mathcal{J}_2,\mathcal{J}_3]=i\mathcal{J}_1,\qquad
[\mathcal{J}_3,\mathcal{J}_1]=i\mathcal{J}_2.
\end{equation}
Introducing the ladder operators $\mathcal{J}_{\pm}=\mathcal{J}_1\pm i\mathcal{J}_2$, the commutation relations become
\begin{equation}
[\mathcal{J}_{3},\mathcal{J}_{\pm}]=\pm \mathcal{J}_{\pm},
\qquad
[\mathcal{J}_{-},\mathcal{J}_{+}]=2\mathcal{J}_{3}.
\label{comm}
\end{equation}
The Casimir operator is an invariant quantity that commutes with all generators of the algebra and therefore characterizes the irreducible representations, playing an essential role in the determination of the physical spectrum. For the $\mathfrak{su}(1,1)$ algebra, the quadratic Casimir operator is
\begin{equation}
\mathcal{C}^2=
\mathcal{J}_3^2-\mathcal{J}_1^2-\mathcal{J}_2^2
=
\mathcal{J}_3^2-\frac{1}{2}
\left(
\mathcal{J}_+\mathcal{J}_-
+
\mathcal{J}_-\mathcal{J}_+
\right),
\qquad
[\mathcal{C}^2,\mathcal{J}_i]=0.
\end{equation}
The $\mathfrak{su}(1,1)$ algebra possesses different series of unitary representations, but here we consider only the discrete series. A basis for an irreducible representation is given by the set $\{|k,n\rangle,\; n=0,1,2,\ldots\}$, where $k$ is the Bargmann index. The action of the generators on these basis states is
\begin{align}
\mathcal{J}_{+}|k,n\rangle
&=
\sqrt{(n+1)(2k+n)}\,|k,n+1\rangle,
\qquad
\mathcal{J}_{-}|k,n\rangle
=
\sqrt{n(2k+n-1)}\,|k,n-1\rangle,
\nonumber\\
\mathcal{J}_{3}|k,n\rangle
&=
(k+n)|k,n\rangle.
\label{basis}
\end{align}
The state $|k,0\rangle$ is the lowest normalized state, while the Casimir operator satisfies $\mathcal{C}^2|k,n\rangle=k(k-1)|k,n\rangle$.

Perelomov coherent states extend the notion of standard coherent states to Lie groups through the action of a displacement operator on a reference state. From a physical point of view, these states describe semiclassical quantum configurations whose dynamics closely resembles the corresponding classical behavior.

For the non-compact group $\mathrm{SU}(1,1)$, the coherent states $|\zeta\rangle$ are defined through the displacement operator
\begin{equation}
\mathcal{D}(\xi)=
\exp(\xi \mathcal{J}_{+}-\xi^{*}\mathcal{J}_{-}),
\qquad
|\zeta\rangle=\mathcal{D}(\xi)|k,0\rangle,
\label{PCS}
\end{equation}
where $\mathcal{J}^{\dagger}_{+}=\mathcal{J}_{-}$ and $\mathcal{J}^{\dagger}_{-}=\mathcal{J}_{+}$ imply
\begin{equation}
\mathcal{D}^{\dagger}(\xi)
=
\exp(\xi^{*}\mathcal{J}_{-}-\xi \mathcal{J}_{+})
=
\mathcal{D}(-\xi).
\label{dis}
\end{equation}
A convenient disentangled form of the displacement operator is
\begin{equation}
\mathcal{D}(\xi)
=
\exp(\zeta \mathcal{J}_{+})
\exp(\eta \mathcal{J}_{3})
\exp(-\zeta^{*}\mathcal{J}_{-}),
\label{normal}
\end{equation}
where $\xi=-\frac{1}{2}\tau e^{-i\varphi}$, $\zeta=-\tanh\left(\frac{1}{2}\tau\right)e^{-i\varphi}$ and $\eta=-2\ln\cosh|\xi|=\ln(1-|\zeta|^2)$ \cite{Gerry}, with $0<\tau<\infty$ and $0\leq\varphi\leq2\pi$. Since $|\zeta|<1$, the coherent parameter evolves inside the unit disk of the complex plane.

Using the normal form of the displacement operator together with Eq.~(\ref{basis}), the Perelomov coherent states can be written as \cite{Perelomov,Perelomov2}
\begin{equation}
|\zeta\rangle
=
(1-|\zeta|^2)^k
\sum_{s=0}^{\infty}
\sqrt{
\frac{\Gamma(s+2k)}
{s!\Gamma(2k)}
}
\,\zeta^s
|k,s\rangle.
\label{PCN}
\end{equation}
When the Hamiltonian belongs to the Lie algebra, the time evolution operator reduces to a group element, and coherent states remain within the same family. This temporal stability transforms the Schr\"odinger equation into classical motion for the coherent parameters. Consider a Hamiltonian proportional to the diagonal generator \cite{Cohen,Kiefer}
\begin{equation}
H=\omega \mathcal{J}_3,
\qquad
U(t)=e^{-iHt}=e^{-i\omega \mathcal{J}_3 t},
\end{equation}
where $\omega$ is a real constant and $\hbar=1$. Using the conjugation properties
\begin{equation}
U(t)\mathcal{J}_{\pm}U^{\dagger}(t)
=
e^{\mp i\omega t}\mathcal{J}_{\pm},
\qquad
U(t)\mathcal{D}(\xi)U^{\dagger}(t)
=
\mathcal{D}(\xi e^{-i\omega t}),
\end{equation}
the time-evolved coherent state becomes
\begin{equation}
|\zeta(t)\rangle
=
U(t)|\zeta\rangle
=
e^{-i\omega kt}\mathcal{D}(\xi e^{-i\omega t})|k,0\rangle.
\end{equation}

Up to an irrelevant global phase, the evolved state remains a Perelomov coherent state whose parameter evolves according to
\begin{equation}
\zeta(t)=\zeta e^{-i\omega t}.
\end{equation}
Thus, under a diagonal Hamiltonian, the coherent state rotates uniformly in the unit disk while preserving its functional form.

\section*{Acknowledgments}
B. C. L. is grateful to the Excellence project FoS UHK 2205/2025-2026 for the financial support.\\
This work was partially supported by SNI--M\'exico, EDI--IPN, and SIP--IPN under project number 20260918.

\section*{Data Availability Statement}
No new data were generated or analyzed in this study.

\section*{ORCID iDs}
M. Salazar-Ram\'irez: \url{https://orcid.org/0000-0003-4139-3026}\\
B. C. L\"utf\"uo\u{g}lu: \url{https://orcid.org/0000-0001-6467-5005}\\
H. Bouguerne: \url{https://orcid.org/0009-0000-0176-4537}

\end{document}